\documentclass[article,twocolumn,preprintnumbers,superscriptaddress,amsmath,amssymb,aps,longbibliography]{revtex4-1}

\usepackage{graphicx}
\usepackage{amssymb,amsmath,hyperref,slashed,color,graphicx,bm,cancel,url,orcidlink}
\usepackage{fontawesome5}
\usepackage[compat=1.1.0]{tikz-feynhand}

\DeclareDocumentCommand{\GitHub}{s}{%
	\IfBooleanTF{#1}{%
		\faGithub~\url{https://github.com/ChitsinYin/Supernova-Graviton-MC}%
	}{%
		\href{https://github.com/ChitsinYin/Supernova-Graviton-MC}{\faGithub}%
	}%
}

\newcommand{\lettersection}[1]{\noindent{\bf \emph{#1}}\,---\,}

\usepackage{catchfilebetweentags}
\usepackage{etoolbox}
\makeatletter
\patchcmd{\CatchFBT@Fin@l}{\endlinechar\m@ne}{}
  {}{\typeout{Unsuccessful patch!}}
\makeatother

\definecolor{niceblue}{rgb}{0.1,0.2,0.6}
\hypersetup{
  colorlinks,
  citecolor = niceblue,
  linkcolor = niceblue,
  urlcolor  = niceblue,
  anchorcolor = niceblue
}

\tikzset{photon/.style={
  thick,
  decorate,
  decoration={snake, amplitude=.45mm, segment length=2mm},
  ->
}}

\tikzset{graviton/.style={
  very thick,
  double,
  decorate,
  decoration={snake, amplitude=.4mm, segment length=1.5mm,
  pre length=.5mm, post length=.5mm},
  ->
}}

\tikzset{fermion/.style={
  thick,
  black,
  postaction={decorate},
  decoration={markings, mark=at position 0.6 with {\arrow{stealth}}}
}}

\newcommand{\dd}{d}
\newcommand{\Rate}{ \frac{\dd\Gamma}{\dd V\,\dd\omega} }
\newcommand{\MP}{M_{\rm P}}

\newcommand{\Li}{\operatorname{Li}}

\begin{document}

\title{At the Highest-Frequency Frontier: Thermal Gravitational Waves from Supernovae}

\author{Xun-Jie Xu\,\orcidlink{0000-0003-3181-1386}}
\email{xuxj@ihep.ac.cn}
\affiliation{Institute of High Energy Physics, Chinese Academy of Sciences, Beijing 100049, China}

\author{Yong Xu\,\orcidlink{0000-0002-4582-8747}}
\email{yong.xu6@mcgill.ca}
\affiliation{McGill University Department of Physics \& Trottier Space Institute, 3600 Rue University, Montréal, QC, H3A 2T8, Canada}

\author{Qiqin Yin\,\orcidlink{0009-0005-7933-3055} }
\email{211870080@smail.nju.edu.cn}
\affiliation{School of Physics, Nanjing University, Nanjing 210093, China}

\date{\today}

\begin{abstract}
Core-collapse supernovae can be powerful sources of high-frequency
gravitational waves (GWs) as the extreme temperatures and densities
in the core allow a considerable number of gravitons to be produced
thermally. We present the first systematic calculation of the thermal
GW spectrum of a core-collapse supernova.  By evaluating the dominant
emission channels---including electron and proton photoproduction
($\gamma e^{-}\to he^{-}$, $\gamma p\to h\,p$) and gravitational bremsstrahlung
($e^{-}p\to e^{-}p\,h$, $e^{-}e^{-}\to e^{-}e^{-}h$)---we find that
a core-collapse supernova generates a total thermal GW luminosity
of $1.27\times10^{26}\ \text{W}$, approximately 18 orders of magnitude
larger than that of the Sun.  The thermal GW spectrum peaks at energies
of $\mathcal{O}(0.1~{\rm GeV})$, extending the predicted spectrum
of known GW sources to an unprecedented high-frequency regime. Our
results provide a benchmark thermal GW spectrum at the high-frequency frontier.  
\end{abstract}
\maketitle


\lettersection{Introduction}Gravitational wave (GW) astronomy has
transformed our view of the Universe through the detection of coherent,
low-frequency signals from compact binary coalescence \cite{LIGOScientific:2016aoc} 
and stochastic backgrounds \cite{NANOGrav:2023gor,EPTA:2023fyk,Reardon:2023gzh,Xu:2023wog}. 
Yet the high- and ultra-high-frequency regimes of the GW spectrum remain a
largely uncharted frontier rich in fundamental physics~\cite{Aggarwal:2020olq,Aggarwal:2025noe}.
Beyond classical motion of macroscopic objects, hot and dense thermal
plasmas in extreme astrophysical and cosmological environments provide
a guaranteed mechanism for generating high-energy gravitons through
microscopic particle collisions. 

It is known that the thermal bath in the early universe can produce
abundant gravitons, which, after cosmic redshift, form a stochastic
GW background~\cite{Ghiglieri:2015nfa,Ghiglieri:2020mhm,Ringwald:2020ist,Bernal:2024jim,Xu:2024cey,Xu:2025wjq}.
Its present-day spectrum typically peaks at ${\cal O}(100)$ GHz,
roughly the same order of magnitude as the CMB peak frequency because
both originate from the same thermal bath and undergo similar redshifts.

Hot stellar media have also been considered for thermal graviton production~\cite{Weinberg:1965nx,1972NCimL...4..927B,Gould1985ApJ,Garcia-Cely:2024ujr}.
More than six decades ago, Weinberg made the first estimate of thermal
gravitons emitted by the Sun, obtaining $6\times10^{7}\text{W}$ for
the solar thermal GW luminosity~\cite{Weinberg:1965nx}. Recently,
the calculation was revisited and extended by Ref.~\cite{Garcia-Cely:2024ujr},
yielding a quantitative spectrum which peaks at $10^{9}$ GHz, with
the total luminosity $1.3\times10^{8}\ \text{W}$. 

\begin{figure}
	\centering
	
	\includegraphics[width=0.99\columnwidth]{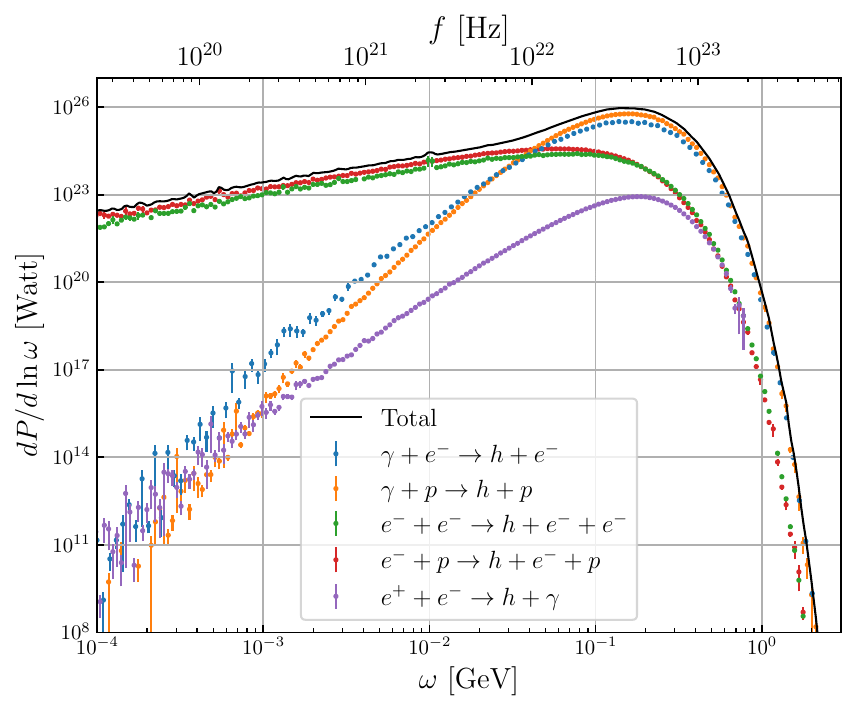}
	
	\caption{The thermal GW power spectrum of a core-collapse supernova obtained
			from Monte Carlo integrations of dominant collision terms.  \label{fig:power-spec}}
\end{figure}

In this work, we push the high-frequency frontier of GWs even further by investigating their thermal production in core-collapse
supernovae. 
We look into the leading thermal channels and find that the supernova-core GW spectrum is dominated by $e^-e^-\to e^-e^-h$ and $e^-p\to e^-ph$ bremsstrahlung at low frequencies, whereas $\gamma e^-\to he^-$ and $\gamma p\to hp$ photoproduction dominates near the spectral peak, as shown in Fig.~\ref{fig:power-spec}. This channel hierarchy marks a qualitative departure from the bremsstrahlung-dominated solar spectrum~\cite{Garcia-Cely:2024ujr}. Combined together,
these production channels generate a total GW luminosity of $1.27\times10^{26}\ \text{W}$,
approximately 18 orders of magnitude higher than the solar GW luminosity. 

%

%
%

Although this work involves similar techniques as those of Ref.~\cite{Garcia-Cely:2024ujr},
we stress that thermal GW production in supernovae requires substantially
more challenging calculations with significantly greater complexity.
First, for processes involving electrons, one must take into account
strong degeneracy effects. The production rate cannot be computed
by simply multiplying relevant cross sections with particle number
densities, an approach often used when Boltzmann statistics is a good
approximation. Instead, one has to compute the rate from more fundamental
collision terms with Pauli blocking effects included.  Second, the
electron gas in a supernova core is relativistic, implying that electrons
cannot be treated as fixed targets in scattering processes and a considerably
high abundance of positrons may also contribute to thermal GW production
via, e.g., $e^{+}e^{-}\to h\gamma$.

\lettersection{Thermal GWs from supernovae} We focus on the post-bounce
phase of a core-collapse supernova, in which a proto-neutron-star
(PNS) core has formed with a typical radius of $10$--$20$ km. 
 The PNS core features extremely high temperatures ($20$--$40$
MeV) and high densities ($10^{14}$--$10^{15}\ \text{g}/\text{cm}^{3}$),
 where frequent particle collisions can efficiently produce gravitons
($h$).  

Let us first consider one of the simplest production channels, $\gamma e^{-}\to he^{-}$,
which also turns out to be one of the dominant production channels.
In this channel, all particles are relativistic. The electrons are
strongly degenerate, with the chemical potential $\mu_{e}$ roughly
5-10 times the temperature $T$. As a consequence, the Pauli-blocking
effect plays a crucial role here, and the production rate of this
channel in supernovae is not a simple rescaling of that in the Sun.
Qualitatively, this channel including the Pauli-blocking effect can
be estimated by considering only electron-hole excitations near the
Fermi surface.  Quantitatively,  the production rate can be computed
as 
\begin{equation}
\Rate=\frac{g_{h}\omega^{2}}{2\pi^{2}}{\cal C}_{h}\thinspace,\label{eq:-1}
\end{equation}
where the production rate $\Rate$ is defined as the number of gravitons
produced per unit volume and per unit graviton energy in the local
medium, $\omega$ denotes the graviton energy, $g_{h}=2$ accounts
for the two graviton polarization states, and ${\cal C}_{h}$ denotes
the collision term for graviton production.  For $\gamma e^{-}\to he^{-}$,
the collision term reads
\begin{equation}
{\cal C}_{h}=\frac{g_{\gamma e}}{2\omega}\int\dd\Pi_{\gamma}\dd\Pi_{e_{1}}\dd\Pi_{e_{2}}(2\pi\delta)^{4}|\overline{{\cal M}}|^{2}f_{\gamma}f_{e_{1}}(1-f_{e_{2}}).\label{eq:}
\end{equation}
Here, $\dd\Pi_{i}=(2\pi)^{-3}\dd^{3}p_{i}/(2E_{i})$ is the Lorentz-invariant
phase-space measure for species $i\in\{\gamma,e_{1},e_{2}\}$, with
$e_{1}$ and $e_{2}$ denoting the initial- and final-state electrons,
respectively. The phase-space distribution of species $i$ is denoted
by $f_{i}$. In addition, $|\overline{{\cal M}}|^{2}$ is the squared
matrix element (with all initial and final spin/polarization averaged)
and $(2\pi\delta)^{4}\equiv(2\pi)^{4}\delta^{(4)}(p_{\gamma}+p_{e_{1}}-p_{e_{2}}-p_{h})$
is the Dirac delta function for energy-momentum conservation. The
factor $g_{\gamma e}=8$ accounts for the internal degrees of freedom
of $\gamma$, $e_{1}$ and $e_{2}$. 

Eq.~\eqref{eq:} is a high-dimensional integral that is generally
difficult to evaluate analytically but can be computed using Monte
Carlo (MC) methods. To perform the MC integration of Eq.~\eqref{eq:},
we employ the open-source {\tt C++} framework {\tt ROOT}~\cite{Brun:1997pa}
as a powerful event generator and develop high-performance numerical
code to compute the integral accurately. Our numerical code is publicly
available on GitHub~\GitHub. 

Analytically, we find that the production rate of $\gamma e^{-}\to he^{-}$
can be approximately estimated by
\begin{equation}
\Rate\simeq\frac{5\alpha G}{2\pi^{3}}\mu_{e}^{2}\frac{\omega^{3}e^{-\omega/T}}{1-e^{-\omega/(2T)}}{\cal F}\left(\frac{\omega}{\kappa}\right).\label{eq:-2}
\end{equation}
Here, $\alpha$ is the fine-structure constant, $G$ is the Newton\textquoteright s
constant, and ${\cal F}\left(\frac{\omega}{\kappa}\right)$ is a function
of $\omega$ and the Debye-H\"uckel screening scale $\kappa$. Its
specific form is ${\cal F}(x)\equiv\theta/\tan\theta$ with $\theta\equiv\arccos(x/2)$
if $x\leq2$, or ${\cal F}(x)\equiv\theta/\tanh\theta$ with $\theta\equiv\text{arccosh}(x/2)$
if $x>2$.  We present a detailed derivation of Eq.~\eqref{eq:-2}
in the Supplemental Material. We  note that, for $\omega\ll T$, Eq.~\eqref{eq:-2} gives $\Rate\propto\omega^3$.
Using $n_{e}\simeq\mu_{e}^{3}/(3\pi^{2})$ in the strongly degenerate
limit, Eq.~\eqref{eq:-2} exhibits an approximate explicit scaling
$\Rate\propto n_{e}^{2/3}$, if the implicit density dependence
through the screening scale $\kappa$ and finite-temperature corrections is neglected.
This differs from the linear $n_{e}$ scaling of solar graviton
photoproduction~\cite{Garcia-Cely:2024ujr}. 

Since the supernova core temperature is well above the electron mass,
positrons ($e^{+}$) are thermally produced frequently and annihilate
with $e^{-}$, implying that gravitons can also be produced via $e^{-}+e^{+}\to h+\gamma$.
This process is closely related to $\gamma+e^{-}\to h+e^{-}$, as
they share essentially the same set of Feynman diagrams. Their squared
matrix elements are related to each other by crossing symmetry and
are therefore of the same order of magnitude. In the early universe
with negligible $\mu_{e}$, the two processes lead to comparable production
rates~\cite{Xu:2025wjq}. In a supernova core, however, the high
chemical potential substantially suppresses the annihilation channel
(the positron-to-electron ratio $n_{e^{+}}/n_{e^{-}}\approx6e^{-\mu_{e}/T}T^{3}/(\mu_{e}^{3}I_{e})$
with $I_{e}=1+\left(\pi T/\mu_{e}\right)^{2}$ is exponentially suppressed
when $\mu_{e}$ is high). Nevertheless, given that $\mu_{e}/T$ is only 
moderately large (hence the exponential suppression is moderate) and
the annihilation channel is free from Pauli blocking in the final
state (unlike the photoproduction channel), it is worth investigating
whether its contribution to the production yield can be significant.
 Hence we also perform a MC analysis for this channel and find that
its yield is about two orders of magnitude lower than that of $\gamma+e^{-}\to h+e^{-}$. 

Similar to $\gamma e^{-}\to he^{-}$, a photon may also scatter off
 a proton to produce a graviton, $\gamma p\to hp$. In this work,
we neglect  photoproduction involving nuclei heavier than a proton
due to their low abundance at such high temperatures.  For $\gamma p\to hp$,
the MC integration is very similar to that of $\gamma e^{-}\to he^{-}$;
while the analytical estimate can be greatly simplified by the nonrelativistic
approximation of protons. Treating them as infinitely-heavy scatters
at rest and assuming $\omega\approx E_{\gamma}$,  one can obtain
a relatively compact analytical formula, as is given in Ref.~\cite{Garcia-Cely:2024ujr}.

Bremsstrahlung processes, such as $e^{-}p\to e^{-}ph$ and $e^{-}e^{-}\to e^{-}e^{-}h$,
play a dominant role in the production of relatively soft gravitons.
Like QED bremsstrahlung, which is closely related to the acceleration/deceleration
of light charged particles, the gravitational bremsstrahlung considered
here is mainly caused by frequent changes in the electron's velocity.
The two processes $e^{-}p\to e^{-}ph$ and $e^{-}e^{-}\to e^{-}e^{-}h$
have comparable production rates with the former being slightly more
efficient than the later (see Fig.~\ref{fig:power-spec} or Tab.~\ref{tab:1}).
Analytically, we obtain the following expressions for their production
rates:
\begin{equation}
\Rate\simeq\frac{2\alpha^{2}G\,\mu_{e}^{2}n_{p}}{\pi^{2}}\frac{1}{e^{\omega/T}-1}\mathcal{K}_{eZ}\left(\frac{\kappa^{2}}{2\mu_{e}^{2}}\right),\label{eq:-3}
\end{equation}
for the $e^{-}p$ channel and 
\begin{equation}
\Rate\simeq\frac{2\alpha^{2}G\mu_{e}^{3}}{3\pi^{5}}\thinspace\frac{4\pi^{2}T^{2}+\omega^{2}}{e^{\omega/T}-1}\mathcal{K}_{ee}\left(\frac{\kappa^{2}}{2\mu_{e}^{2}}\right),\label{eq:-23}
\end{equation}
for the $e^{-}e^{-}$ channel, with $\mathcal{K}_{eZ}\left(x\right)\simeq2\ln^{2}\left(\frac{2}{x}\right)-\frac{10}{3}\ln\left(\frac{2}{x}\right)+\frac{2\pi^{2}}{3}-\frac{16}{3}$
and $\mathcal{K}_{ee}\left(x\right)\simeq\frac{39}{4}-\frac{7}{2}\ln x+\frac{8}{3}\ln^{2}x$.
In the strongly degenerate limit, Eq.~\eqref{eq:-3} exhibits an
approximate explicit scaling $n_{e}^{2/3}n_{p}$, while Eq.~\eqref{eq:-23}
exhibits an explicit scaling $n_{e}$ rather than $n_{e}^{2}$. These
scalings are understood at fixed $T$ and $\omega$, apart from the
implicit dependence of the screening functions on the plasma parameters. In the regime with $\omega \ll T$,  bremsstrahlung graviton production admits a scaling $\Rate \propto T/\omega$.

%
%

Given a specific supernova profile, one can integrate the local production
rates above throughout the supernova core to obtain the GW emission
power. We adopt a benchmark supernova profile simulated by the Garching
group~\cite{Bollig:2020xdr,Mirizzi:2015eza,garching} and perform
the integration as follows:
\begin{equation}
\frac{\dd P}{\dd\omega}\equiv\int\dd^{3}{\bf r}\,\omega\frac{\dd\Gamma({\bf r})}{\dd V\,\dd\omega}\thinspace.\label{eq:coll_power_general}
\end{equation}
The result is presented in Fig.~\ref{fig:power-spec}, where the
colored points with error bars represent the MC integrations of several
dominant channels, and the black line represents the total GW power
spectrum. 

As shown in Fig.~\ref{fig:power-spec}, the total thermal GW power
spectrum peaks around $\omega\sim0.15 \text{ GeV}$, reaching a maximum
luminosity of $\mathcal{O}(10^{26})\text{ W}$. The high-frequency
regime ($\omega\gtrsim0.04\text{ GeV}$) is dominated by photoproduction
processes ($\gamma e^{-}\to he^{-}$ and $\gamma p\to hp$). In contrast,
the low-frequency regime ($\omega\lesssim0.04\text{ GeV}$)  is driven
by gravitational bremsstrahlung processes ($e^{-}p\to e^{-}ph$ and
$e^{-}e^{-}\to e^{-}e^{-}h$). 

\begin{figure}
\centering

\includegraphics[width=0.99\columnwidth]{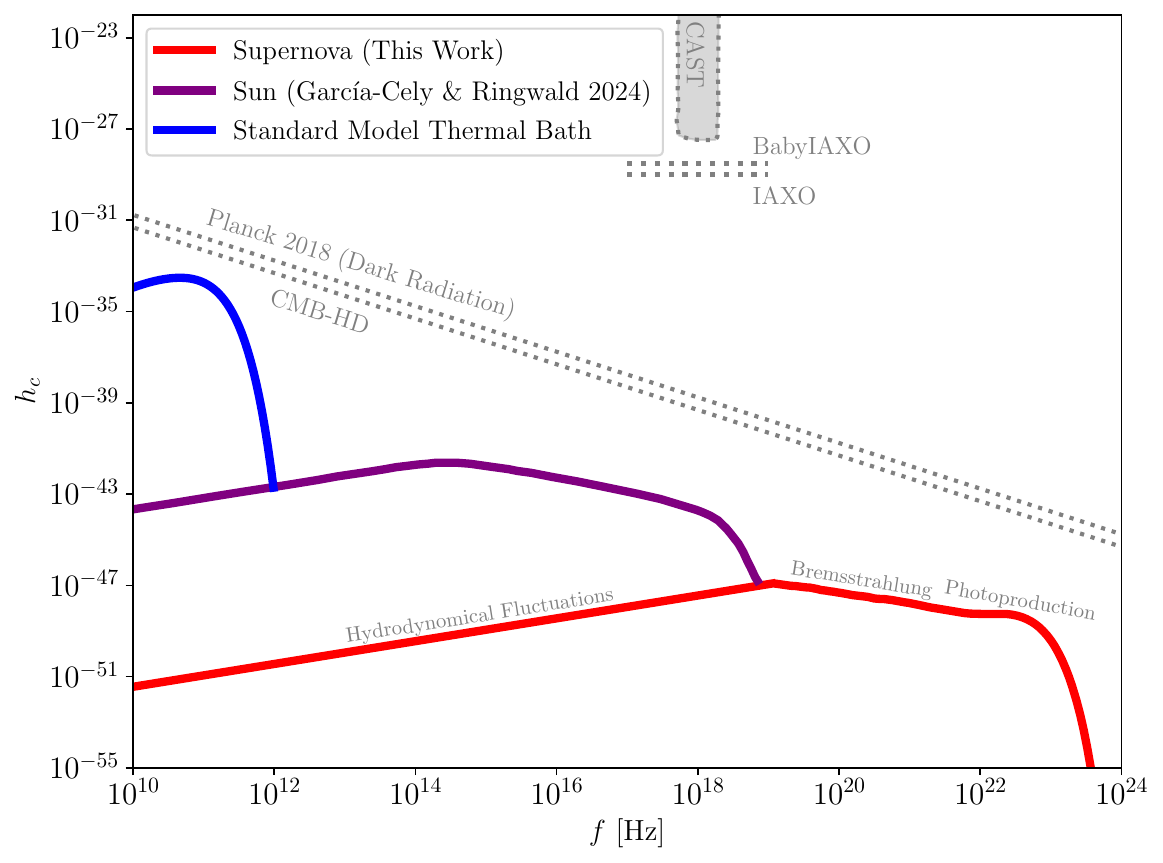}

\caption{Characteristic GW strain $h_c(f)$ as observed at Earth. Red: microscopic GW spectrum from a core-collapse supernova located at a distance of $10~{\rm kpc}$. Purple: thermal GW spectrum from the Sun at the Earth’s location. Blue: reference cosmological GW spectrum from a Standard Model thermal bath, evaluated at the present epoch.}
\label{fig:hc}
\end{figure}

Before closing this section, we comment on  two additional mechanisms
for low-frequency GW production in supernovae. The first arises from
thermal fluctuations of the energy-momentum tensor at macroscopic
scales. At length scales above the mean-free-path of microscopic particles,
individual particle collisions are no longer a valid description and
the supernova core behaves as a viscous fluid. In this regime, GWs
can still be generated thermally through fluctuation of the anisotropic
stress tensor, with the power spectrum given by~\cite{Ghiglieri:2015nfa,Garcia-Cely:2024ujr}
\begin{equation}
\frac{\dd P}{\dd\omega}=\frac{16G\omega^{2}}{\pi}\int\dd^{3}r\,\eta(r)T(r)\thinspace,\label{eq:hydro_power}
\end{equation}
where $\eta$ is the shear viscosity, which can be computed using
 the prescription in Ref.~\cite{PhysRevE.90.033105}. The calculation
of this part of contribution is very similar to that in Ref.~\cite{Garcia-Cely:2024ujr},
yielding $\dd P/\dd\omega\propto\omega^{2}$ or $\dd P/\dd\ln\omega\propto\omega^{3}$
at low frequencies; and the result is included in Fig.~\ref{fig:hc}.
 We note that in terms of the integrated radiation power, the contribution
of Eq.~\eqref{eq:hydro_power} is negligibly small compared to that
of microscopic particle collisions. 

At even larger length scales relevant to the shock dynamics such as
standing accretion shock instabilities (SASI), convection, and rotation,
complex hydrodynamic processes during supernova evolution can generate
significant GWs below the kilohertz scale. This topic has been actively
investigated using dedicated supernova simulations; see, e.g., Refs.~\cite{Ott:2008wt,Abdikamalov:2020jzn,Vartanyan:2023sxm,Mezzacappa:2024zph,Lella:2026aat}.

\emph{}

\lettersection{Radiation power  and characteristic strain} Integrating
the full spectrum gives the total GW radiation power of a supernova
core, 
\begin{equation}
P_{{\rm tot}}=\int\dd\omega\,\frac{\dd P}{\dd\omega}\simeq1.27\times10^{26}~{\rm W}\thinspace.\label{eq:total_power}
\end{equation}
Here, the contributions from multiple relevant GW production channels
are summed. The obtained value is approximately $10^{18}$ times larger
than the solar GW radiation power, $1.3\times10^{8}~{\rm W}$ \cite{Garcia-Cely:2024ujr}.
In Tab.~\ref{tab:1}, we list the contribution of each channel. In
terms of the radiation power, $\gamma p\to hp$ is the most productive
channel. About $87\%$ of the total power is generated by photoproduction
processes. In terms of the number of gravitons emitted, however, bremsstrahlung
processes are more efficient: about $64\%$ gravitons are produced
by $e^{-}p\to e^{-}ph$ and $e^{-}e^{-}\to e^{-}e^{-}h$. 

\begin{table}

\begin{tabular}{cccc}
\hline 
Channel & $P$ & \phantom{xx}$P/P_{\text{tot}}$\phantom{xx} & \phantom{xx}$N/N_{\text{tot}}$\phantom{xx}\tabularnewline
\hline 
\phantom{xx} $\gamma e^{-}\to he^{-}$\phantom{xx} & $4.0\times10^{25}$ W & $31.6\%$ & $13.5\%$\tabularnewline
$\gamma p\to hp$ & $7.1\times10^{25}$ W & $55.8\%$ & $22.8\%$\tabularnewline
$e^{-}p\to e^{-}ph$ & $9.5\times10^{24}$ W & $7.4\%$ & $39.0\%$\tabularnewline
$e^{-}e^{-}\to e^{-}e^{-}h$  & $6.5\times10^{24}$ W & $5.1\%$ & $24.6\%$\tabularnewline
$e^{+}e^{-}\to h\gamma$ & $9.9\times10^{22}$ W & $0.07\%$ & $0.03\%$\tabularnewline
\hline 
\end{tabular}\caption{\label{tab:1} Graviton radiation power $P$, its fraction of the
total power $P/P_{\text{tot}}$, and its fraction of the total graviton
number emission rate $N/N_{\text{tot}}$ for each production channel
in a typical supernova core.}

\end{table}

For a Galactic supernova at a characteristic distance $d_{{\rm SN}}=10~{\rm kpc}$,
 the differential graviton flux at Earth is given by 
\begin{equation}
\frac{\dd\Phi}{\dd\omega}=\frac{1}{4\pi d_{{\rm SN}}^{2}}\frac{1}{\omega}\frac{\dd P}{\dd\omega}\thinspace.\label{eq:differential_flux}
\end{equation}
Integrating over frequency yields the total flux:
\begin{equation}
\Phi_{{\rm tot}}=\int\dd\omega\,\frac{\dd\Phi}{\dd\omega}\simeq13~{\rm km^{-2}\,s^{-1}}\thinspace.\label{eq:sn_flux}
\end{equation}
Since a core-collapse supernova generally lasts a few seconds, Eq.~\eqref{eq:sn_flux}
implies that a Galactic event typically deposit ${\cal O}(10^{9})$
high-energy gravitons on Earth. 

For high-frequency GWs, the characteristic strain $h_{c}$ provides
a useful measure of the signal strength. It is related to the differential
power spectrum by 
\begin{equation}
h_{c}(\omega)=\frac{2}{d_{{\rm SN}}}\sqrt{\frac{G}{\omega}\frac{\dd P}{\dd\omega}}\thinspace.\label{eq:characteristic_strain}
\end{equation}
 Figure~\ref{fig:hc} displays $h_{c}$ calculated for a core-collapse
supernova at $d_{\text{SN}} = 10~{\rm kpc}$ alongside those for other high-frequency
GW sources including the Sun~\cite{Garcia-Cely:2024ujr} and the
Standard Model thermal bath in the early universe~\cite{Ghiglieri:2015nfa,Ghiglieri:2020mhm,Ringwald:2020ist}. We also show constraints from dark radiation, including the current Planck 2018 bound \cite{Planck:2018vyg} and the projected sensitivity of CMB-HD \cite{CMB-HD:2022bsz}, indicated by gray dotted lines. The gray shaded region represents the constraint derived from CAST  \cite{Ejlli:2019bqj,CAST:2017uph}, while the two dashed lines indicate the projected sensitivities of IAXO and BabyIAXO \cite{IAXO:2019mpb}. The blue  line corresponds to the thermal GWs produced in early universe, whose amplitude depends on the reheating
temperature, which is set at $T_{{\rm rh}}\simeq5.5\times10^{15}~{\rm GeV}$.
 The red curve shows the total supernona GW spectrum. At the lowest frequencies, the emission is dominated by hydrodynamic fluctuations, for which Eq.~\eqref{eq:hydro_power} gives $h_c\propto f^{1/2}$. Our main new results are the microscopic contributions, obtained by summing the five channels shown in Fig.~\ref{fig:power-spec}. Among these channels, bremsstrahlung dominates at graviton energies $\omega\lesssim 0.04~\mathrm{GeV}$ ($f\lesssim 10^{22}~\mathrm{Hz}$), yielding $h_c\propto f^{-1/2}$. At higher frequencies, photoproduction becomes dominant and drives  approximately $h_c\propto f^{3/2}$ toward the spectral peak at $\omega\simeq 0.15~\mathrm{GeV}$ ($f\simeq 4\times 10^{22}~\mathrm{Hz}$), beyond which the spectrum is exponentially suppressed. This behavior differs qualitatively from the solar GW spectrum, where bremsstrahlung remains dominant: in the SN core, photoproduction controls the emission near the spectral maximum, producing a pronounced high-frequency peak and a characteristic change in spectral slope.
  As shown in the figure, the SN spectrum occupies a substantially higher-frequency band than the solar and cosmological spectra. Despite its enormous  graviton luminosity, the strain observed at Earth is reduced by geometric dilution over the much larger source distance compared with the Sun.

\lettersection{Observations via the Gertsenshtein effect} 
 Like other high-frequency GWs shown in Fig.~\ref{fig:hc}, supernova
thermal GWs represent a formidable observational challenge.  To
date, no proposed high-frequency GW detector has reached the sensitivity
required to detect these signals~\cite{Aggarwal:2020olq,Aggarwal:2025noe}.
Nevertheless, we discuss a scenario which, albeit highly optimistic,
might be of phenomenological interest.

When a graviton traverses a magnetic field, it can be converted into
a photon via the Gertsenshtein effect~\cite{gertsenshtein_wave_1962,Boccaletti:1970pxw}.
For a  magnetic field  perpendicular to the graviton momentum, the
conversion probability reads:
\begin{equation}
P_{h\to\gamma}=\sin^{2}\left(\frac{B\ell}{\sqrt{2}M_{P}}\right)\simeq3.3\times10^{-7}\left(\frac{B}{10^{11}\text{T}}\thinspace\frac{\ell}{20\ \text{km}}\right)^{2},\label{eq:-4}
\end{equation}
where $B$ denotes the magnetic field strength, $\ell$ is the propagation
length, and $M_{P}$ is the reduced Planck mass. The strongest magnetic
fields in the Universe are generated by a class of neutron stars known
as \textit{magnetars}, reaching up to $10^{11}$ T~\cite{Kaspi:2017fwg}.
The surface dipolar field of a magnetar can maintain this magnitude
over an extent of roughly $20\ \text{km}$. According to Eq.~\eqref{eq:-4},
if gravitons produced by a supernova pass through this field before
reaching Earth, approximately a fraction of $10^{-7}$ will be converted
into photons. Multiplying this conversion rate by the aforementioned
total number of arriving gravitons, ${\cal O}(10^{9})$, we obtain
${\cal O}(10^{2})$ photons deposited on Earth. Although detecting
such photons would require an extremely large exposure areas and exceptionally
low background levels, it is not entirely impossible.

\lettersection{Summary and conclusions} More than six decades ago,
Weinberg showed that ordinary stellar environments such as the Sun
continuously emit gravitational radiation from stochastic microscopic
processes in a thermal plasma \cite{Weinberg:1965nx}, an idea recently
revisited using modern techniques \cite{Garcia-Cely:2024ujr}. In this \emph{Letter}, we extend this microscopic framework to the extreme environment of a core-collapse supernova and present the first systematic calculation of its thermal graviton spectrum.

By treating the supernova core as a hot, dense, relativistic, and
degenerate electron-proton plasma, we have systematically evaluated
the ultra-high-frequency thermal GW emission. Incorporating full quantum
statistics---specifically Pauli blocking for electrons---we evaluated
the dominant reaction channels, including electron and proton photoproduction
($\gamma e^{-}\to he^{-}$, $\gamma p\to hp$) and gravitational bremsstrahlung
($e^{-}p\to e^{-}ph$, $e^{-}e^{-}\to e^{-}e^{-}h$). Our  calculations
show that photoproduction and bremsstrahlung dominate the hard and
soft portions of the thermal GW spectrum, respectively. The total
integrated GW luminosity of a typical supernova core is estimated
to be $P_{{\rm tot}}\simeq1.27\times10^{26}~{\rm W}$, which exceeds
the solar thermal graviton luminosity by approximately 18 orders of
magnitude.
The contribution of each production channel is listed in Tab.~\ref{tab:1}, which shows that photoproduction dominates the luminosity ($87\%$), while bremsstrahlung contributes the largest fraction of the emitted gravitons ($64\%$).

Regarding observability, we find that a Galactic supernova would 
deliver a total number of ${\cal O}(10^{9})$ thermal gravitons to
Earth. Although this number is too small for direct detection by any
proposed terrestrial detectors, we propose a highly optimistic, yet
not entirely impossible, scenario in which the ${\cal O}(10^{9})$
gravitons might be of phenomenological significance.

Our results establish supernova cores as novel astrophysical sources
of ultra-high-frequency GWs and provide a concrete benchmark for future
detection efforts. 

\begin{acknowledgments}
We are grateful to Camilo García-Cely and Andreas Ringwald for useful discussions.
This work is supported in
part by the National Natural Science Foundation of China under grant
No.~12141501 and also by the CAS Project for Young Scientists in
Basic Research (YSBR-099). YX is supported by the Natural Sciences and Engineering Research Council (NSERC) of Canada.
\end{acknowledgments}

\bibliographystyle{apsrev4-1}
\bibliography{biblio}

\begin{thebibliography}{36}%
\makeatletter
\providecommand \@ifxundefined [1]{%
 \@ifx{#1\undefined}
}%
\providecommand \@ifnum [1]{%
 \ifnum #1\expandafter \@firstoftwo
 \else \expandafter \@secondoftwo
 \fi
}%
\providecommand \@ifx [1]{%
 \ifx #1\expandafter \@firstoftwo
 \else \expandafter \@secondoftwo
 \fi
}%
\providecommand \natexlab [1]{#1}%
\providecommand \enquote  [1]{``#1''}%
\providecommand \bibnamefont  [1]{#1}%
\providecommand \bibfnamefont [1]{#1}%
\providecommand \citenamefont [1]{#1}%
\providecommand \href@noop [0]{\@secondoftwo}%
\providecommand \href [0]{\begingroup \@sanitize@url \@href}%
\providecommand \@href[1]{\@@startlink{#1}\@@href}%
\providecommand \@@href[1]{\endgroup#1\@@endlink}%
\providecommand \@sanitize@url [0]{\catcode `\\12\catcode `\$12\catcode
  `\&12\catcode `\#12\catcode `\^12\catcode `\_12\catcode `\%12\relax}%
\providecommand \@@startlink[1]{}%
\providecommand \@@endlink[0]{}%
\providecommand \url  [0]{\begingroup\@sanitize@url \@url }%
\providecommand \@url [1]{\endgroup\@href {#1}{\urlprefix }}%
\providecommand \urlprefix  [0]{URL }%
\providecommand \Eprint [0]{\href }%
\providecommand \doibase [0]{http://dx.doi.org/}%
\providecommand \selectlanguage [0]{\@gobble}%
\providecommand \bibinfo  [0]{\@secondoftwo}%
\providecommand \bibfield  [0]{\@secondoftwo}%
\providecommand \translation [1]{[#1]}%
\providecommand \BibitemOpen [0]{}%
\providecommand \bibitemStop [0]{}%
\providecommand \bibitemNoStop [0]{.\EOS\space}%
\providecommand \EOS [0]{\spacefactor3000\relax}%
\providecommand \BibitemShut  [1]{\csname bibitem#1\endcsname}%
\let\auto@bib@innerbib\@empty
\bibitem [{\citenamefont {Abbott}\ \emph {et~al.}(2016)\citenamefont {Abbott}
  \emph {et~al.}}]{LIGOScientific:2016aoc}%
  \BibitemOpen
  \bibfield  {author} {\bibinfo {author} {\bibfnamefont {B.~P.}\ \bibnamefont
  {Abbott}} \emph {et~al.} (\bibinfo {collaboration} {LIGO Scientific,
  Virgo}),\ }\href {\doibase 10.1103/PhysRevLett.116.061102} {\bibfield
  {journal} {\bibinfo  {journal} {Phys. Rev. Lett.}\ }\textbf {\bibinfo
  {volume} {116}},\ \bibinfo {pages} {061102} (\bibinfo {year} {2016})},\
  \Eprint {http://arxiv.org/abs/1602.03837} {arXiv:1602.03837 [gr-qc]}
  \BibitemShut {NoStop}%
\bibitem [{\citenamefont {Agazie}\ \emph {et~al.}(2023)\citenamefont {Agazie}
  \emph {et~al.}}]{NANOGrav:2023gor}%
  \BibitemOpen
  \bibfield  {author} {\bibinfo {author} {\bibfnamefont {G.}~\bibnamefont
  {Agazie}} \emph {et~al.} (\bibinfo {collaboration} {NANOGrav}),\ }\href
  {\doibase 10.3847/2041-8213/acdac6} {\bibfield  {journal} {\bibinfo
  {journal} {Astrophys. J. Lett.}\ }\textbf {\bibinfo {volume} {951}},\
  \bibinfo {pages} {L8} (\bibinfo {year} {2023})},\ \Eprint
  {http://arxiv.org/abs/2306.16213} {arXiv:2306.16213 [astro-ph.HE]}
  \BibitemShut {NoStop}%
\bibitem [{\citenamefont {Antoniadis}\ \emph {et~al.}(2023)\citenamefont
  {Antoniadis} \emph {et~al.}}]{EPTA:2023fyk}%
  \BibitemOpen
  \bibfield  {author} {\bibinfo {author} {\bibfnamefont {J.}~\bibnamefont
  {Antoniadis}} \emph {et~al.} (\bibinfo {collaboration} {EPTA, InPTA}),\
  }\href {\doibase 10.1051/0004-6361/202346844} {\bibfield  {journal} {\bibinfo
   {journal} {Astron. Astrophys.}\ }\textbf {\bibinfo {volume} {678}},\
  \bibinfo {pages} {A50} (\bibinfo {year} {2023})},\ \Eprint
  {http://arxiv.org/abs/2306.16214} {arXiv:2306.16214 [astro-ph.HE]}
  \BibitemShut {NoStop}%
\bibitem [{\citenamefont {Reardon}\ \emph {et~al.}(2023)\citenamefont {Reardon}
  \emph {et~al.}}]{Reardon:2023gzh}%
  \BibitemOpen
  \bibfield  {author} {\bibinfo {author} {\bibfnamefont {D.~J.}\ \bibnamefont
  {Reardon}} \emph {et~al.},\ }\href {\doibase 10.3847/2041-8213/acdd02}
  {\bibfield  {journal} {\bibinfo  {journal} {Astrophys. J. Lett.}\ }\textbf
  {\bibinfo {volume} {951}},\ \bibinfo {pages} {L6} (\bibinfo {year} {2023})},\
  \Eprint {http://arxiv.org/abs/2306.16215} {arXiv:2306.16215 [astro-ph.HE]}
  \BibitemShut {NoStop}%
\bibitem [{\citenamefont {Xu}\ \emph {et~al.}(2023)\citenamefont {Xu} \emph
  {et~al.}}]{Xu:2023wog}%
  \BibitemOpen
  \bibfield  {author} {\bibinfo {author} {\bibfnamefont {H.}~\bibnamefont {Xu}}
  \emph {et~al.},\ }\href {\doibase 10.1088/1674-4527/acdfa5} {\bibfield
  {journal} {\bibinfo  {journal} {Res. Astron. Astrophys.}\ }\textbf {\bibinfo
  {volume} {23}},\ \bibinfo {pages} {075024} (\bibinfo {year} {2023})},\
  \Eprint {http://arxiv.org/abs/2306.16216} {arXiv:2306.16216 [astro-ph.HE]}
  \BibitemShut {NoStop}%
\bibitem [{\citenamefont {Aggarwal}\ \emph {et~al.}(2021)\citenamefont
  {Aggarwal} \emph {et~al.}}]{Aggarwal:2020olq}%
  \BibitemOpen
  \bibfield  {author} {\bibinfo {author} {\bibfnamefont {N.}~\bibnamefont
  {Aggarwal}} \emph {et~al.},\ }\href {\doibase 10.1007/s41114-021-00032-5}
  {\bibfield  {journal} {\bibinfo  {journal} {Living Rev. Rel.}\ }\textbf
  {\bibinfo {volume} {24}},\ \bibinfo {pages} {4} (\bibinfo {year} {2021})},\
  \Eprint {http://arxiv.org/abs/2011.12414} {arXiv:2011.12414 [gr-qc]}
  \BibitemShut {NoStop}%
\bibitem [{\citenamefont {Aggarwal}\ \emph {et~al.}(2025)\citenamefont
  {Aggarwal} \emph {et~al.}}]{Aggarwal:2025noe}%
  \BibitemOpen
  \bibfield  {author} {\bibinfo {author} {\bibfnamefont {N.}~\bibnamefont
  {Aggarwal}} \emph {et~al.},\ }\href@noop {} {\  (\bibinfo {year} {2025})},\
  \Eprint {http://arxiv.org/abs/2501.11723} {arXiv:2501.11723 [gr-qc]}
  \BibitemShut {NoStop}%
\bibitem [{\citenamefont {Ghiglieri}\ and\ \citenamefont
  {Laine}(2015)}]{Ghiglieri:2015nfa}%
  \BibitemOpen
  \bibfield  {author} {\bibinfo {author} {\bibfnamefont {J.}~\bibnamefont
  {Ghiglieri}}\ and\ \bibinfo {author} {\bibfnamefont {M.}~\bibnamefont
  {Laine}},\ }\href {\doibase 10.1088/1475-7516/2015/07/022} {\bibfield
  {journal} {\bibinfo  {journal} {JCAP}\ }\textbf {\bibinfo {volume} {07}},\
  \bibinfo {pages} {022} (\bibinfo {year} {2015})},\ \Eprint
  {http://arxiv.org/abs/1504.02569} {arXiv:1504.02569 [hep-ph]} \BibitemShut
  {NoStop}%
\bibitem [{\citenamefont {Ghiglieri}\ \emph {et~al.}(2020)\citenamefont
  {Ghiglieri}, \citenamefont {Jackson}, \citenamefont {Laine},\ and\
  \citenamefont {Zhu}}]{Ghiglieri:2020mhm}%
  \BibitemOpen
  \bibfield  {author} {\bibinfo {author} {\bibfnamefont {J.}~\bibnamefont
  {Ghiglieri}}, \bibinfo {author} {\bibfnamefont {G.}~\bibnamefont {Jackson}},
  \bibinfo {author} {\bibfnamefont {M.}~\bibnamefont {Laine}}, \ and\ \bibinfo
  {author} {\bibfnamefont {Y.}~\bibnamefont {Zhu}},\ }\href {\doibase
  10.1007/JHEP07(2020)092} {\bibfield  {journal} {\bibinfo  {journal} {JHEP}\
  }\textbf {\bibinfo {volume} {07}},\ \bibinfo {pages} {092} (\bibinfo {year}
  {2020})},\ \Eprint {http://arxiv.org/abs/2004.11392} {arXiv:2004.11392
  [hep-ph]} \BibitemShut {NoStop}%
\bibitem [{\citenamefont {Ringwald}\ \emph {et~al.}(2021)\citenamefont
  {Ringwald}, \citenamefont {Sch{\"u}tte-Engel},\ and\ \citenamefont
  {Tamarit}}]{Ringwald:2020ist}%
  \BibitemOpen
  \bibfield  {author} {\bibinfo {author} {\bibfnamefont {A.}~\bibnamefont
  {Ringwald}}, \bibinfo {author} {\bibfnamefont {J.}~\bibnamefont
  {Sch{\"u}tte-Engel}}, \ and\ \bibinfo {author} {\bibfnamefont
  {C.}~\bibnamefont {Tamarit}},\ }\href {\doibase
  10.1088/1475-7516/2021/03/054} {\bibfield  {journal} {\bibinfo  {journal}
  {JCAP}\ }\textbf {\bibinfo {volume} {03}},\ \bibinfo {pages} {054} (\bibinfo
  {year} {2021})},\ \Eprint {http://arxiv.org/abs/2011.04731} {arXiv:2011.04731
  [hep-ph]} \BibitemShut {NoStop}%
\bibitem [{\citenamefont {Bernal}\ and\ \citenamefont
  {Xu}(2025)}]{Bernal:2024jim}%
  \BibitemOpen
  \bibfield  {author} {\bibinfo {author} {\bibfnamefont {N.}~\bibnamefont
  {Bernal}}\ and\ \bibinfo {author} {\bibfnamefont {Y.}~\bibnamefont {Xu}},\
  }\href {\doibase 10.1007/JHEP01(2025)137} {\bibfield  {journal} {\bibinfo
  {journal} {JHEP}\ }\textbf {\bibinfo {volume} {01}},\ \bibinfo {pages} {137}
  (\bibinfo {year} {2025})},\ \Eprint {http://arxiv.org/abs/2410.21385}
  {arXiv:2410.21385 [hep-ph]} \BibitemShut {NoStop}%
\bibitem [{\citenamefont {Xu}(2025)}]{Xu:2024cey}%
  \BibitemOpen
  \bibfield  {author} {\bibinfo {author} {\bibfnamefont {Y.}~\bibnamefont
  {Xu}},\ }\href {\doibase 10.1016/j.physletb.2025.139483} {\bibfield
  {journal} {\bibinfo  {journal} {Phys. Lett. B}\ }\textbf {\bibinfo {volume}
  {865}},\ \bibinfo {pages} {139483} (\bibinfo {year} {2025})},\ \Eprint
  {http://arxiv.org/abs/2412.21137} {arXiv:2412.21137 [hep-ph]} \BibitemShut
  {NoStop}%
\bibitem [{\citenamefont {Xu}\ \emph {et~al.}(2025)\citenamefont {Xu},
  \citenamefont {Xu}, \citenamefont {Yin},\ and\ \citenamefont
  {Zhu}}]{Xu:2025wjq}%
  \BibitemOpen
  \bibfield  {author} {\bibinfo {author} {\bibfnamefont {X.-J.}\ \bibnamefont
  {Xu}}, \bibinfo {author} {\bibfnamefont {Y.}~\bibnamefont {Xu}}, \bibinfo
  {author} {\bibfnamefont {Q.}~\bibnamefont {Yin}}, \ and\ \bibinfo {author}
  {\bibfnamefont {J.}~\bibnamefont {Zhu}},\ }\href {\doibase
  10.1007/JHEP10(2025)141} {\bibfield  {journal} {\bibinfo  {journal} {JHEP}\
  }\textbf {\bibinfo {volume} {10}},\ \bibinfo {pages} {141} (\bibinfo {year}
  {2025})},\ \Eprint {http://arxiv.org/abs/2505.08868} {arXiv:2505.08868
  [hep-ph]} \BibitemShut {NoStop}%
\bibitem [{\citenamefont {Weinberg}(1965)}]{Weinberg:1965nx}%
  \BibitemOpen
  \bibfield  {author} {\bibinfo {author} {\bibfnamefont {S.}~\bibnamefont
  {Weinberg}},\ }\href {\doibase 10.1103/PhysRev.140.B516} {\bibfield
  {journal} {\bibinfo  {journal} {Phys. Rev.}\ }\textbf {\bibinfo {volume}
  {140}},\ \bibinfo {pages} {B516} (\bibinfo {year} {1965})}\BibitemShut
  {NoStop}%
\bibitem [{\citenamefont {{Boccaletti}}(1972)}]{1972NCimL...4..927B}%
  \BibitemOpen
  \bibfield  {author} {\bibinfo {author} {\bibfnamefont {D.}~\bibnamefont
  {{Boccaletti}}},\ }\href@noop {} {\bibfield  {journal} {\bibinfo  {journal}
  {Nuovo Cimento Lettere}\ }\textbf {\bibinfo {volume} {4}},\ \bibinfo {pages}
  {927} (\bibinfo {year} {1972})}\BibitemShut {NoStop}%
\bibitem [{\citenamefont {{Gould}}(1985)}]{Gould1985ApJ}%
  \BibitemOpen
  \bibfield  {author} {\bibinfo {author} {\bibfnamefont {R.~J.}\ \bibnamefont
  {{Gould}}},\ }\href {\doibase 10.1086/162848} {\bibfield  {journal} {\bibinfo
   {journal} {\apj}\ }\textbf {\bibinfo {volume} {288}},\ \bibinfo {pages}
  {789} (\bibinfo {year} {1985})}\BibitemShut {NoStop}%
\bibitem [{\citenamefont {Garc{\'\i}a-Cely}\ and\ \citenamefont
  {Ringwald}(2025)}]{Garcia-Cely:2024ujr}%
  \BibitemOpen
  \bibfield  {author} {\bibinfo {author} {\bibfnamefont {C.}~\bibnamefont
  {Garc{\'\i}a-Cely}}\ and\ \bibinfo {author} {\bibfnamefont {A.}~\bibnamefont
  {Ringwald}},\ }\href {\doibase 10.1103/gtwg-pr41} {\bibfield  {journal}
  {\bibinfo  {journal} {Phys. Rev. Lett.}\ }\textbf {\bibinfo {volume} {135}},\
  \bibinfo {pages} {061001} (\bibinfo {year} {2025})},\ \Eprint
  {http://arxiv.org/abs/2407.18297} {arXiv:2407.18297 [hep-ph]} \BibitemShut
  {NoStop}%
\bibitem [{\citenamefont {Brun}\ and\ \citenamefont
  {Rademakers}(1997)}]{Brun:1997pa}%
  \BibitemOpen
  \bibfield  {author} {\bibinfo {author} {\bibfnamefont {R.}~\bibnamefont
  {Brun}}\ and\ \bibinfo {author} {\bibfnamefont {F.}~\bibnamefont
  {Rademakers}},\ }\href {\doibase 10.1016/s0168-9002(97)00048-x} {\bibfield
  {journal} {\bibinfo  {journal} {Nucl. Instrum. Meth. A}\ }\textbf {\bibinfo
  {volume} {389}},\ \bibinfo {pages} {81} (\bibinfo {year} {1997})}\BibitemShut
  {NoStop}%
\bibitem [{\citenamefont {Bollig}\ \emph {et~al.}(2020)\citenamefont {Bollig},
  \citenamefont {DeRocco}, \citenamefont {Graham},\ and\ \citenamefont
  {Janka}}]{Bollig:2020xdr}%
  \BibitemOpen
  \bibfield  {author} {\bibinfo {author} {\bibfnamefont {R.}~\bibnamefont
  {Bollig}}, \bibinfo {author} {\bibfnamefont {W.}~\bibnamefont {DeRocco}},
  \bibinfo {author} {\bibfnamefont {P.~W.}\ \bibnamefont {Graham}}, \ and\
  \bibinfo {author} {\bibfnamefont {H.-T.}\ \bibnamefont {Janka}},\ }\href
  {\doibase 10.1103/PhysRevLett.125.051104} {\bibfield  {journal} {\bibinfo
  {journal} {Phys. Rev. Lett.}\ }\textbf {\bibinfo {volume} {125}},\ \bibinfo
  {pages} {051104} (\bibinfo {year} {2020})},\ \bibinfo {note} {[Erratum:
  Phys.Rev.Lett. 126, 189901 (2021)]},\ \Eprint
  {http://arxiv.org/abs/2005.07141} {arXiv:2005.07141 [hep-ph]} \BibitemShut
  {NoStop}%
\bibitem [{\citenamefont {Mirizzi}\ \emph {et~al.}(2016)\citenamefont
  {Mirizzi}, \citenamefont {Tamborra}, \citenamefont {Janka}, \citenamefont
  {Saviano}, \citenamefont {Scholberg}, \citenamefont {Bollig}, \citenamefont
  {Hudepohl},\ and\ \citenamefont {Chakraborty}}]{Mirizzi:2015eza}%
  \BibitemOpen
  \bibfield  {author} {\bibinfo {author} {\bibfnamefont {A.}~\bibnamefont
  {Mirizzi}}, \bibinfo {author} {\bibfnamefont {I.}~\bibnamefont {Tamborra}},
  \bibinfo {author} {\bibfnamefont {H.-T.}\ \bibnamefont {Janka}}, \bibinfo
  {author} {\bibfnamefont {N.}~\bibnamefont {Saviano}}, \bibinfo {author}
  {\bibfnamefont {K.}~\bibnamefont {Scholberg}}, \bibinfo {author}
  {\bibfnamefont {R.}~\bibnamefont {Bollig}}, \bibinfo {author} {\bibfnamefont
  {L.}~\bibnamefont {Hudepohl}}, \ and\ \bibinfo {author} {\bibfnamefont
  {S.}~\bibnamefont {Chakraborty}},\ }\href {\doibase
  10.1393/ncr/i2016-10120-8} {\bibfield  {journal} {\bibinfo  {journal} {Riv.
  Nuovo Cim.}\ }\textbf {\bibinfo {volume} {39}},\ \bibinfo {pages} {1}
  (\bibinfo {year} {2016})},\ \Eprint {http://arxiv.org/abs/1508.00785}
  {arXiv:1508.00785 [astro-ph.HE]} \BibitemShut {NoStop}%
\bibitem [{gar()}]{garching}%
  \BibitemOpen
  \href@noop {} {}\bibinfo {howpublished} {The Garching Core-Collapse Supernova
  Archive, \url{https://wwwmpa.mpa-garching.mpg.de/ccsnarchive/}. The data set
  used in our work is taken from ``Hydro radial profiles'' in ``Data and
  supplementary material of 1D simulations with PROMETHEUS-VERTEX''. We select
  the 18.80 solar-mass progenitor in this data set as the benchmark
  profile.}\BibitemShut {Stop}%
\bibitem [{\citenamefont {Daligault}\ \emph {et~al.}(2014)\citenamefont
  {Daligault}, \citenamefont {Rasmussen},\ and\ \citenamefont
  {Baalrud}}]{PhysRevE.90.033105}%
  \BibitemOpen
  \bibfield  {author} {\bibinfo {author} {\bibfnamefont {J.}~\bibnamefont
  {Daligault}}, \bibinfo {author} {\bibfnamefont {K.~O.}\ \bibnamefont
  {Rasmussen}}, \ and\ \bibinfo {author} {\bibfnamefont {S.~D.}\ \bibnamefont
  {Baalrud}},\ }\href {\doibase 10.1103/PhysRevE.90.033105} {\bibfield
  {journal} {\bibinfo  {journal} {Phys. Rev. E}\ }\textbf {\bibinfo {volume}
  {90}},\ \bibinfo {pages} {033105} (\bibinfo {year} {2014})}\BibitemShut
  {NoStop}%
\bibitem [{\citenamefont {Ott}(2009)}]{Ott:2008wt}%
  \BibitemOpen
  \bibfield  {author} {\bibinfo {author} {\bibfnamefont {C.}~\bibnamefont
  {Ott}},\ }\href {\doibase 10.1088/0264-9381/26/6/063001} {\bibfield
  {journal} {\bibinfo  {journal} {Class. Quant. Grav.}\ }\textbf {\bibinfo
  {volume} {26}},\ \bibinfo {pages} {063001} (\bibinfo {year} {2009})},\
  \Eprint {http://arxiv.org/abs/0809.0695} {arXiv:0809.0695 [astro-ph]}
  \BibitemShut {NoStop}%
\bibitem [{\citenamefont {Abdikamalov}\ \emph {et~al.}(2020)\citenamefont
  {Abdikamalov}, \citenamefont {Pagliaroli},\ and\ \citenamefont
  {Radice}}]{Abdikamalov:2020jzn}%
  \BibitemOpen
  \bibfield  {author} {\bibinfo {author} {\bibfnamefont {E.}~\bibnamefont
  {Abdikamalov}}, \bibinfo {author} {\bibfnamefont {G.}~\bibnamefont
  {Pagliaroli}}, \ and\ \bibinfo {author} {\bibfnamefont {D.}~\bibnamefont
  {Radice}},\ }\href {\doibase 10.1007/978-981-15-4702-7\_21-1} {\  (\bibinfo
  {year} {2020}),\ 10.1007/978-981-15-4702-7\_21-1},\ \Eprint
  {http://arxiv.org/abs/2010.04356} {arXiv:2010.04356 [astro-ph.SR]}
  \BibitemShut {NoStop}%
\bibitem [{\citenamefont {Vartanyan}\ \emph {et~al.}(2023)\citenamefont
  {Vartanyan}, \citenamefont {Burrows}, \citenamefont {Wang}, \citenamefont
  {Coleman},\ and\ \citenamefont {White}}]{Vartanyan:2023sxm}%
  \BibitemOpen
  \bibfield  {author} {\bibinfo {author} {\bibfnamefont {D.}~\bibnamefont
  {Vartanyan}}, \bibinfo {author} {\bibfnamefont {A.}~\bibnamefont {Burrows}},
  \bibinfo {author} {\bibfnamefont {T.}~\bibnamefont {Wang}}, \bibinfo {author}
  {\bibfnamefont {M.~S.~B.}\ \bibnamefont {Coleman}}, \ and\ \bibinfo {author}
  {\bibfnamefont {C.~J.}\ \bibnamefont {White}},\ }\href {\doibase
  10.1103/PhysRevD.107.103015} {\bibfield  {journal} {\bibinfo  {journal}
  {Phys. Rev. D}\ }\textbf {\bibinfo {volume} {107}},\ \bibinfo {pages}
  {103015} (\bibinfo {year} {2023})},\ \Eprint
  {http://arxiv.org/abs/2302.07092} {arXiv:2302.07092 [astro-ph.HE]}
  \BibitemShut {NoStop}%
\bibitem [{\citenamefont {Mezzacappa}\ and\ \citenamefont
  {Zanolin}(2024)}]{Mezzacappa:2024zph}%
  \BibitemOpen
  \bibfield  {author} {\bibinfo {author} {\bibfnamefont {A.}~\bibnamefont
  {Mezzacappa}}\ and\ \bibinfo {author} {\bibfnamefont {M.}~\bibnamefont
  {Zanolin}},\ }\href@noop {} {\  (\bibinfo {year} {2024})},\ \Eprint
  {http://arxiv.org/abs/2401.11635} {arXiv:2401.11635 [astro-ph.HE]}
  \BibitemShut {NoStop}%
\bibitem [{\citenamefont {Lella}\ \emph {et~al.}(2026)\citenamefont {Lella},
  \citenamefont {Lucente}, \citenamefont {Kresse}, \citenamefont {Glas},
  \citenamefont {Janka},\ and\ \citenamefont {Mirizzi}}]{Lella:2026aat}%
  \BibitemOpen
  \bibfield  {author} {\bibinfo {author} {\bibfnamefont {A.}~\bibnamefont
  {Lella}}, \bibinfo {author} {\bibfnamefont {G.}~\bibnamefont {Lucente}},
  \bibinfo {author} {\bibfnamefont {D.}~\bibnamefont {Kresse}}, \bibinfo
  {author} {\bibfnamefont {R.}~\bibnamefont {Glas}}, \bibinfo {author}
  {\bibfnamefont {H.-T.}\ \bibnamefont {Janka}}, \ and\ \bibinfo {author}
  {\bibfnamefont {A.}~\bibnamefont {Mirizzi}},\ }\href {\doibase
  10.1103/f3n4-k2cq} {\bibfield  {journal} {\bibinfo  {journal} {Phys. Rev. D}\
  }\textbf {\bibinfo {volume} {113}},\ \bibinfo {pages} {083034} (\bibinfo
  {year} {2026})},\ \Eprint {http://arxiv.org/abs/2602.02651} {arXiv:2602.02651
  [astro-ph.HE]} \BibitemShut {NoStop}%
\bibitem [{\citenamefont {Aghanim}\ \emph {et~al.}(2020)\citenamefont {Aghanim}
  \emph {et~al.}}]{Planck:2018vyg}%
  \BibitemOpen
  \bibfield  {author} {\bibinfo {author} {\bibfnamefont {N.}~\bibnamefont
  {Aghanim}} \emph {et~al.} (\bibinfo {collaboration} {Planck}),\ }\href
  {\doibase 10.1051/0004-6361/201833910} {\bibfield  {journal} {\bibinfo
  {journal} {Astron. Astrophys.}\ }\textbf {\bibinfo {volume} {641}},\ \bibinfo
  {pages} {A6} (\bibinfo {year} {2020})},\ \bibinfo {note} {[Erratum:
  Astron.Astrophys. 652, C4 (2021)]},\ \Eprint
  {http://arxiv.org/abs/1807.06209} {arXiv:1807.06209 [astro-ph.CO]}
  \BibitemShut {NoStop}%
\bibitem [{\citenamefont {Aiola}\ \emph {et~al.}(2022)\citenamefont {Aiola}
  \emph {et~al.}}]{CMB-HD:2022bsz}%
  \BibitemOpen
  \bibfield  {author} {\bibinfo {author} {\bibfnamefont {S.}~\bibnamefont
  {Aiola}} \emph {et~al.} (\bibinfo {collaboration} {CMB-HD}),\ }\href@noop {}
  {\  (\bibinfo {year} {2022})},\ \Eprint {http://arxiv.org/abs/2203.05728}
  {arXiv:2203.05728 [astro-ph.CO]} \BibitemShut {NoStop}%
\bibitem [{\citenamefont {Ejlli}\ \emph {et~al.}(2019)\citenamefont {Ejlli},
  \citenamefont {Ejlli}, \citenamefont {Cruise}, \citenamefont {Pisano},\ and\
  \citenamefont {Grote}}]{Ejlli:2019bqj}%
  \BibitemOpen
  \bibfield  {author} {\bibinfo {author} {\bibfnamefont {A.}~\bibnamefont
  {Ejlli}}, \bibinfo {author} {\bibfnamefont {D.}~\bibnamefont {Ejlli}},
  \bibinfo {author} {\bibfnamefont {A.~M.}\ \bibnamefont {Cruise}}, \bibinfo
  {author} {\bibfnamefont {G.}~\bibnamefont {Pisano}}, \ and\ \bibinfo {author}
  {\bibfnamefont {H.}~\bibnamefont {Grote}},\ }\href {\doibase
  10.1140/epjc/s10052-019-7542-5} {\bibfield  {journal} {\bibinfo  {journal}
  {Eur. Phys. J. C}\ }\textbf {\bibinfo {volume} {79}},\ \bibinfo {pages}
  {1032} (\bibinfo {year} {2019})},\ \Eprint {http://arxiv.org/abs/1908.00232}
  {arXiv:1908.00232 [gr-qc]} \BibitemShut {NoStop}%
\bibitem [{\citenamefont {Anastassopoulos}\ \emph {et~al.}(2017)\citenamefont
  {Anastassopoulos} \emph {et~al.}}]{CAST:2017uph}%
  \BibitemOpen
  \bibfield  {author} {\bibinfo {author} {\bibfnamefont {V.}~\bibnamefont
  {Anastassopoulos}} \emph {et~al.} (\bibinfo {collaboration} {CAST}),\ }\href
  {\doibase 10.1038/nphys4109} {\bibfield  {journal} {\bibinfo  {journal}
  {Nature Phys.}\ }\textbf {\bibinfo {volume} {13}},\ \bibinfo {pages} {584}
  (\bibinfo {year} {2017})},\ \Eprint {http://arxiv.org/abs/1705.02290}
  {arXiv:1705.02290 [hep-ex]} \BibitemShut {NoStop}%
\bibitem [{\citenamefont {Armengaud}\ \emph {et~al.}(2019)\citenamefont
  {Armengaud} \emph {et~al.}}]{IAXO:2019mpb}%
  \BibitemOpen
  \bibfield  {author} {\bibinfo {author} {\bibfnamefont {E.}~\bibnamefont
  {Armengaud}} \emph {et~al.} (\bibinfo {collaboration} {IAXO}),\ }\href
  {\doibase 10.1088/1475-7516/2019/06/047} {\bibfield  {journal} {\bibinfo
  {journal} {JCAP}\ }\textbf {\bibinfo {volume} {06}},\ \bibinfo {pages} {047}
  (\bibinfo {year} {2019})},\ \Eprint {http://arxiv.org/abs/1904.09155}
  {arXiv:1904.09155 [hep-ph]} \BibitemShut {NoStop}%
\bibitem [{\citenamefont {Ade}\ \emph {et~al.}(2021)\citenamefont {Ade} \emph
  {et~al.}}]{BICEP:2021xfz}%
  \BibitemOpen
  \bibfield  {author} {\bibinfo {author} {\bibfnamefont {P.~A.~R.}\
  \bibnamefont {Ade}} \emph {et~al.} (\bibinfo {collaboration} {BICEP, Keck}),\
  }\href {\doibase 10.1103/PhysRevLett.127.151301} {\bibfield  {journal}
  {\bibinfo  {journal} {Phys. Rev. Lett.}\ }\textbf {\bibinfo {volume} {127}},\
  \bibinfo {pages} {151301} (\bibinfo {year} {2021})},\ \Eprint
  {http://arxiv.org/abs/2110.00483} {arXiv:2110.00483 [astro-ph.CO]}
  \BibitemShut {NoStop}%
\bibitem [{\citenamefont {Gertsenshtein}(1962)}]{gertsenshtein_wave_1962}%
  \BibitemOpen
  \bibfield  {author} {\bibinfo {author} {\bibfnamefont {M.}~\bibnamefont
  {Gertsenshtein}},\ }\href@noop {} {\bibfield  {journal} {\bibinfo  {journal}
  {Sov. Phys. JETP}\ }\textbf {\bibinfo {volume} {14}},\ \bibinfo {pages} {84}
  (\bibinfo {year} {1962})},\ \bibinfo {note}
  {\url{http://jetp.ras.ru/cgi-bin/r/index/r/41/1/p113?a=list}}\BibitemShut
  {NoStop}%
\bibitem [{\citenamefont {Boccaletti}\ \emph {et~al.}(1970)\citenamefont
  {Boccaletti}, \citenamefont {De~Sabbata}, \citenamefont {Fortini},\ and\
  \citenamefont {Gualdi}}]{Boccaletti:1970pxw}%
  \BibitemOpen
  \bibfield  {author} {\bibinfo {author} {\bibfnamefont {D.}~\bibnamefont
  {Boccaletti}}, \bibinfo {author} {\bibfnamefont {V.}~\bibnamefont
  {De~Sabbata}}, \bibinfo {author} {\bibfnamefont {P.}~\bibnamefont {Fortini}},
  \ and\ \bibinfo {author} {\bibfnamefont {C.}~\bibnamefont {Gualdi}},\ }\href
  {\doibase 10.1007/BF02710177} {\bibfield  {journal} {\bibinfo  {journal}
  {Nuovo Cim. B}\ }\textbf {\bibinfo {volume} {70}},\ \bibinfo {pages} {129}
  (\bibinfo {year} {1970})}\BibitemShut {NoStop}%
\bibitem [{\citenamefont {Kaspi}\ and\ \citenamefont
  {Beloborodov}(2017)}]{Kaspi:2017fwg}%
  \BibitemOpen
  \bibfield  {author} {\bibinfo {author} {\bibfnamefont {V.~M.}\ \bibnamefont
  {Kaspi}}\ and\ \bibinfo {author} {\bibfnamefont {A.}~\bibnamefont
  {Beloborodov}},\ }\href {\doibase 10.1146/annurev-astro-081915-023329}
  {\bibfield  {journal} {\bibinfo  {journal} {Ann. Rev. Astron. Astrophys.}\
  }\textbf {\bibinfo {volume} {55}},\ \bibinfo {pages} {261} (\bibinfo {year}
  {2017})},\ \Eprint {http://arxiv.org/abs/1703.00068} {arXiv:1703.00068
  [astro-ph.HE]} \BibitemShut {NoStop}%
\end{thebibliography}%

\clearpage

\onecolumngrid

\setcounter{equation}{0}
\setcounter{figure}{0}
\setcounter{table}{0}
\setcounter{page}{1}
\setcounter{section}{0}

\makeatletter
\renewcommand{\thesection}{S\arabic{section}}
\renewcommand{\theequation}{S\arabic{equation}}
\renewcommand{\thefigure}{S\arabic{figure}}
\renewcommand{\thetable}{S\arabic{table}}

\renewcommand{\theHfigure}{S\arabic{figure}}
\renewcommand{\theHtable}{S\arabic{table}}
\renewcommand{\theHequation}{S\arabic{equation}}
\makeatother

\begin{center}
    \textbf{\large At the Highest-Frequency Frontier: Thermal Gravitational Waves from Supernovae} \\ 
    \vspace{0.5cm}
    { \it \large Supplemental Material}\\ 
    \vspace{0.5cm}
    {Xun-Jie Xu, Yong Xu, Qiqin Yin}
\end{center}

\begin{figure}[h]
\centering

\includegraphics[width=0.99\columnwidth]{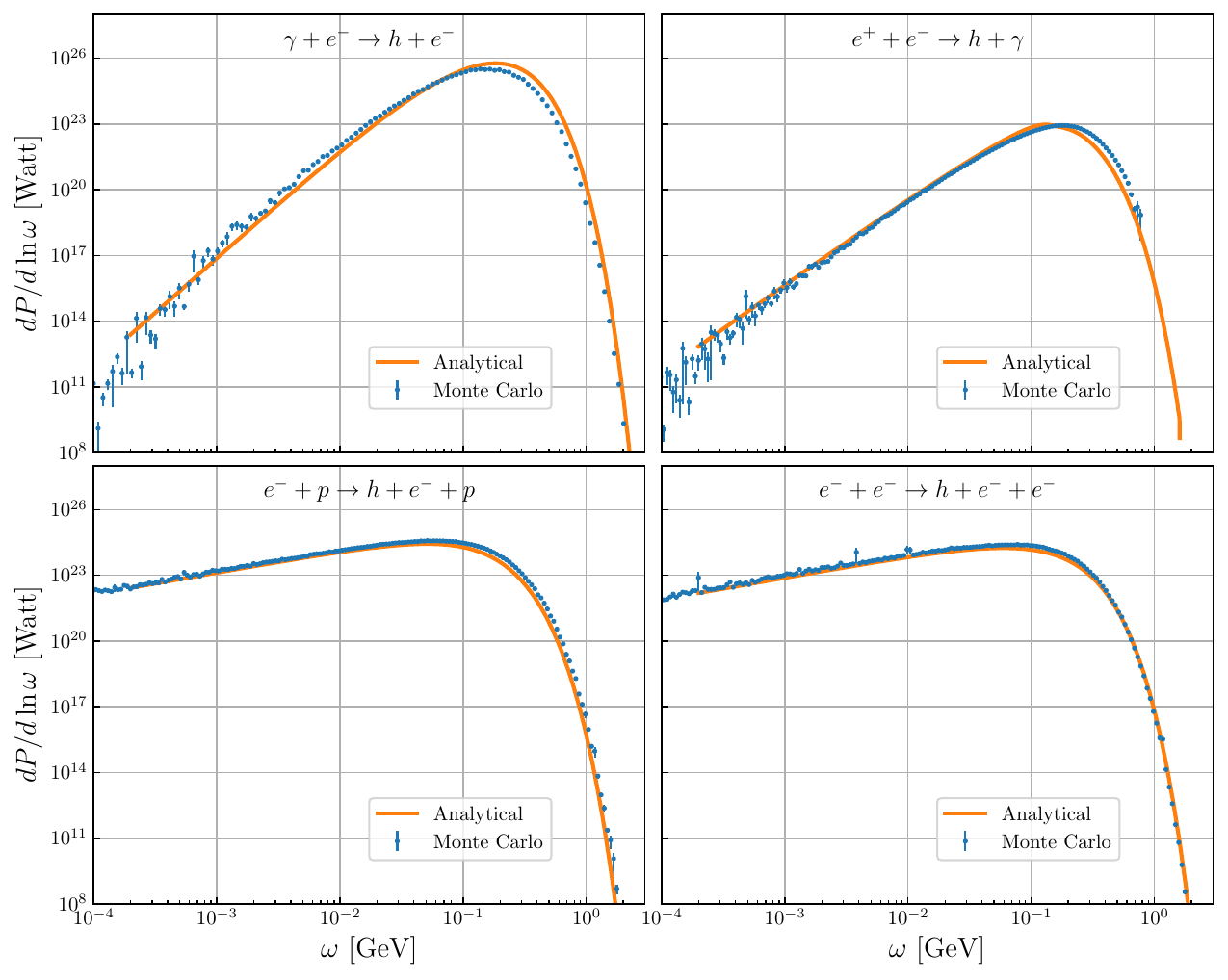}

\caption{Analytical (orange lines) and MC results (blue points) of supernova
thermal GW power spectra, calculated for $\gamma e^{-}\to he^{-}$,
$e^{+}e^{-}\to h\gamma$, $e^{-}e^{-}\to e^{-}e^{-}h$, and $e^{-}p\to e^{-}ph$.
\label{fig:MC-ana}}
\end{figure}

This Supplemental Material analytically derives the thermal production
rates of gravitons in relativistic and degenerate electron-proton
plasma. The analytical results are checked against direct Monte Carlo
(MC) integration of the corresponding collision terms.  In Fig.~\ref{fig:MC-ana},
we present both analytical and MC results of the GW power spectra
for $\gamma e^{-}\to he^{-}$, $e^{+}e^{-}\to h\gamma$, $e^{-}e^{-}\to e^{-}e^{-}h$,
and $e^{-}p\to e^{-}ph$. As is shown in the figure, our analytical
results reach reasonably good agreement with the MC results. The comparison
indicates that for semi-quantitative estimate of supernova thermal
GW production rates, the analytical formulae presented in this work
should  suffice.  Our code of MC integration is publicly available
on GitHub~\GitHub.

Below we present the derivation in details. We use natural units $\hbar=c=k_{B}=1$,
metric signature $(+,-,-,-)$, $8\pi M_{P}^{2}=1/G$, and $\alpha=e^{2}/(4\pi)$.
The four-momentum of a emitted graviton is denoted by $k^{\mu}=(\omega,{\bf k})$,
with $|{\bf k}|=\omega$.  The electron, photon, and proton number
densities are denoted by $n_{e}$, $n_{p}$, and $n_{\gamma}$, respectively;
and the chemical potential of electrons is denoted by $\mu_{e}$.
When both electrons and positrons are present in an analysis, we use
$n_{e^{-}}$ and $n_{e^{+}}$ to denote their number densities. 

\section{Useful formulae for degenerate electron gas}

\begin{figure}[h]
\centering

\includegraphics[width=0.8\columnwidth]{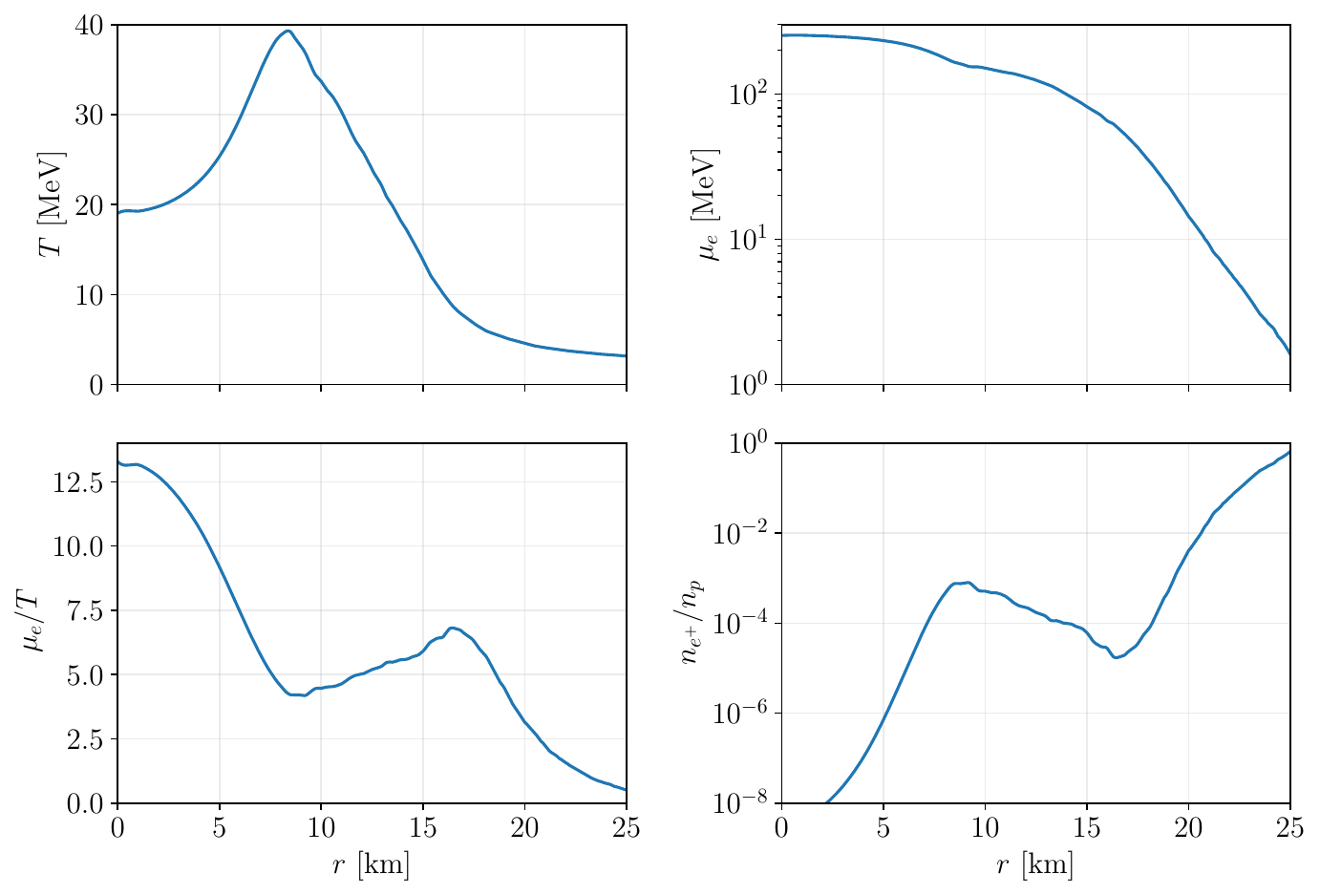}

\caption{\label{fig:profile}The core profile of a supernova. Data taken from
the Garching supernova simulation~\cite{garching}. }
\end{figure}

In a supernova core, electrons can be approximately treated as degenerate,
relativistic fermion gas, with the following phase-space distribution
function
\begin{equation}
f_{e}(p)=\frac{1}{\exp\left(\frac{p-\mu_{e}}{T}\right)+1}\thinspace.\label{eq:-27}
\end{equation}
 If the degeneracy is sufficiently strong, $f_{e}(p)$ can be approximated
as a step function: $f_{e}=\Theta(\mu_{e}-p)$. In this limit, the
electron number density is given by 
\begin{equation}
n_{e}^{(0)}\equiv\frac{g_{e}}{(2\pi)^{3}}\int_{0}^{\mu_{e}}4\pi p^{2}dp=\frac{g_{e}}{2}\thinspace\frac{\mu_{e}^{3}}{3\pi^{2}}\thinspace\ \ \ \ (\text{for}\ \mu_{e}/T\gg1)\thinspace.\label{eq:-28}
\end{equation}
Since the typical value of the chemical potential $\mu_{e}$ is only
several times larger than $T$ (see Fig.~\ref{fig:profile}, where
we present the core profile of a supernova simulated by the Garching
group~\cite{Bollig:2020xdr,Mirizzi:2015eza,garching}), the deviation
of $n_{e}$ from the above strong degeneracy limit can be significant.
 To account for this significant deviation, we define 
\begin{equation}
I_{e}\equiv\frac{n_{e}}{n_{e}^{(0)}}\thinspace.\label{eq:-29}
\end{equation}
With $n_{e}=\frac{g_{e}}{(2\pi)^{3}}\int d^{3}pf_{e}(p)$ evaluated
at small $T$, one obtains the finite-$T$ correction below 
\begin{equation}
I_{e}\approx1+\left(\frac{\pi T}{\mu_{e}}\right)^{2}.\label{eq:-30}
\end{equation}
From the $\mu_{e}/T$ ratio presented in Fig.~\ref{fig:profile},
we see that the second term in Eq.~\eqref{eq:-30} may reach ${\cal O}(1)$
and is not fully negligible. Given that  the above ${\cal O}(T^{2}/\mu_{e}^{2})$
correction can be significant, one may be concerned with higher order
corrections. It turns out that they are actually significantly smaller
than the $T^{2}/\mu_{e}^{2}$ correction at $T/\mu_{e}={\cal O}(1)$.
For instance, when $\mu_{e}/T\in\{3,\ 2,\ 1\}$, the true value of
$I_{e}$ is $\{2.10762,\ 3.56725,\ 12.985\}$, to be compared with
$\{2.09662,\ 3.4674,\ 10.8696\}$ from Eq.~\eqref{eq:-30}. 

Flipping the sign of $\mu_{e}$ in Eq.~\eqref{eq:-27} gives the phase-space
distribution function of positrons ($e^{+}$). The number density
of $e^{+}$ can be approximately computed using the Boltzmann approximation:
\begin{equation}
n_{e^{+}}\approx g_{e}\frac{T^{3}}{\pi^{2}}e^{-\mu_{e}/T}\thinspace.\label{eq:-31}
\end{equation}
Comparing it with $n_{e^{-}}=n_{e}=n_{e}^{(0)}I_{e}$, we obtain
\begin{equation}
\frac{n_{e^{+}}}{n_{e^{-}}}\approx6\frac{T^{3}}{\mu_{e}^{3}I_{e}}e^{-\mu_{e}/T}\thinspace.\label{eq:-32}
\end{equation}

Electric charge neutrality requires 
\begin{equation}
n_{e^{-}}-n_{e^{+}}\approx n_{p}\thinspace,\label{eq:-33}
\end{equation}
which is not an exact relation due to the presence of a small number
of other charged particles (such as muons) in the  supernova core. 

Substituting Eq.~\eqref{eq:-31} and $n_{e^{-}}=n_{e}^{(0)}I_{e}$
into Eq.~\eqref{eq:-33}, we obtain
\begin{equation}
\mu_{e}^{3}+\pi^{2}T^{2}\mu_{e}-6T^{3}e^{-\mu_{e}/T}-3\pi^{2}n_{p}\approx0\thinspace,\label{eq:-34}
\end{equation}
which can be used to determine $\mu_{e}$ from $T$ and $n_{p}$. 

\section{Photoproduction: $e^{-}+\gamma\to e^{-}+h$ \label{sec:e-gamma}}

The momenta of initial and final states in this process are  denoted
as follows
\begin{equation}
e^{-}(p_{1})+\gamma(p_{2})\to e^{-}(k_{1})+h(k)\thinspace.\label{eq:photo_process}
\end{equation}
According to Eqs.~\eqref{eq:-1} and \eqref{eq:}, the graviton production
rate for this process reads
\begin{equation}
\Rate={\cal N}_{\gamma e}\int\frac{\dd^{3}p_{1}}{2E_{p_{1}}}\thinspace\frac{\dd^{3}p_{2}}{2E_{p_{2}}}\thinspace\frac{\dd^{3}k_{1}}{2E_{k_{1}}}\thinspace\frac{\omega^{2}\dd\Omega_{h}}{2\omega}f_{\gamma}(p_{2})f_{e}(p_{1})[1-f_{e}(k_{1})]|\overline{\mathcal{M}}|^{2}(2\pi)^{4}\delta^{(4)}(p_{1}+p_{2}-k_{1}-k)\thinspace,\label{eq:-5}
\end{equation}
with
\begin{equation}
{\cal N}_{\gamma e}=\frac{g_{e}^{2}g_{\gamma}g_{h}}{(2\pi)^{12}}\thinspace,\label{eq:-6}
\end{equation}
where $g_{e}=g_{\gamma}=g_{h}=2$ account for the internal degrees
of freedom of electrons, photons, and gravitons, respectively.

Note that $|\overline{\mathcal{M}}|^{2}$ is the squared matrix element
with both initial- and final-state spins/polarizations averaged, while
in particle scattering the squared matrix element $|{\cal M}|^{2}$
with final-state spins/polarizations summed. They are related to each
other by
\begin{equation}
|{\cal M}|^{2}=g_{e}g_{h}|\overline{\mathcal{M}}|^{2}\thinspace.\label{eq:-7}
\end{equation}
We proceed with $|{\cal M}|^{2}$ and rewrite Eq.~\eqref{eq:-5} as
follows:
\begin{align}
\Rate= & \frac{n_{\gamma}n_{e}}{(2\pi)^{6}N_{\gamma e}}\omega\int\frac{\dd^{3}p_{1}\dd^{3}p_{2}}{4E_{p_{1}}E_{p_{2}}}\frac{\dd^{3}k_{1}\dd\Omega_{h}}{4E_{k_{1}}}f_{\gamma}(p_{2})f_{e}(p_{1})[1-f_{e}(k_{1})]|{\cal M}|^{2}(2\pi\delta)^{4}\thinspace,\label{eq:photo_rate_start}
\end{align}
where 
\begin{equation}
N_{\gamma e}\equiv g_{e}g_{h}\times\frac{n_{\gamma}n_{e}}{(2\pi)^{6}{\cal N}_{\gamma e}}=\int\dd^{3}p_{2}\,f_{\gamma}(p_{2})\int\dd^{3}p_{1}f_{e}(p_{1})\thinspace.\label{eq:-8}
\end{equation}

 In the analytic estimate below we use $f_{\gamma}(p_{2})\simeq e^{-p_{2}/T}$
(our numerical integration still retains the full Bose-Einstein distribution
for the photon) and take the degenerate and relativistic limit for
the electron. In this limit,    the $N_{\gamma e}$ factor is
approximated as
\begin{equation}
N_{\gamma e}\simeq\frac{32\pi^{2}T^{3}\mu_{e}^{3}}{3}I_{e}\thinspace.\label{eq:-9}
\end{equation}

Energy conservation gives $p_{1}+p_{2}=k_{1}+\omega$, and hence 
\begin{align}
f_{\gamma}(p_{2})f_{e}(p_{1})[1-f_{e}(k_{1})] & =e^{-p_{2}/T}\frac{1}{e^{(p_{1}-\mu)/T}+1}\frac{e^{(k_{1}-\mu)/T}}{e^{(k_{1}-\mu)/T}+1}=e^{-\omega/T}[1-f_{e}(p_{1})]f_{e}(k_{1})\thinspace.\label{eq:photo_stat_identity}
\end{align}

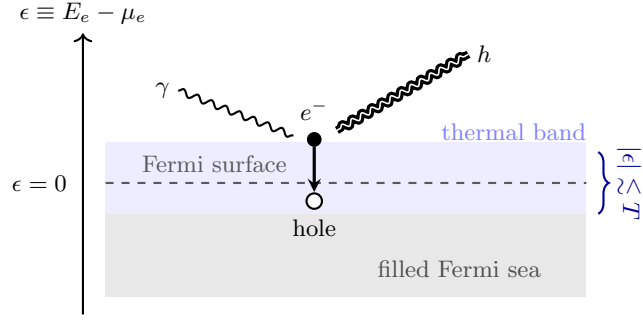
\begin{figure}[t]
\centering \begin{tikzpicture}[scale=1.2]

\tikzset{photon/.style={
  thick,
  decorate,
  decoration={snake, amplitude=.42mm, segment length=2mm}
}}

\tikzset{graviton/.style={
  very thick,
  double,
  decorate,
  decoration={snake, amplitude=.35mm, segment length=1.5mm,
  pre length=.5mm, post length=.5mm}
}}

\tikzset{fermion/.style={
  thick,
  black,
  ->,
  >=stealth
}}

\draw[->, thick] (-2.9,-1.45) -- (-2.9,1.65)
  node[above] {$\epsilon\equiv E_e-\mu_e$};

\fill[gray!18] (-2.65,-1.25) rectangle (2.65,0);

\fill[blue!7] (-2.65,-0.34) rectangle (2.65,0.45);

\draw[black!65, dashed, thick] (-2.65,0) -- (2.65,0);
\node[left] at (-2.98,0) {$\epsilon=0$};
\node[black!65, above] at (-1.45,0.03) {Fermi surface};

\draw[
  decorate,
  decoration={brace, amplitude=4pt},
  blue!60!black,
  thick
]
  (2.78,0.34) -- (2.78,-0.34);

\node[blue!60!black, rotate=-90] at (3.12,0)
  {$|\epsilon| \lesssim T$};
\node[blue!50, above] at (1.85,0.38) {thermal band}; 
\node[gray!60!black] at (1.25,-0.95) {filled Fermi sea};

\filldraw[black] (-0.35,0.48) circle (0.075);
\node[above] at (-0.35,0.60) {$e^-$};

\filldraw[white, draw=black, thick] (-0.35,-0.20) circle (0.085);
\node[below] at (-0.35,-0.31) {hole};

\draw[fermion, very thick] (-0.35,0.40) -- (-0.35,-0.10);

\draw[photon] (-1.85,1.02) -- (-0.58,0.52);
\node[left] at (-1.85,1.02) {$\gamma$};

\draw[graviton] (-0.10,0.58) -- (1.35,1.42);
\node[right] at (1.35,1.42) {$h$};

\end{tikzpicture}

\caption{Schematic of photoproduction in a degenerate supernova core. The vertical
coordinate is the electron excitation energy $\epsilon=E_{e}-\mu_{e}$.
The dashed line $\epsilon=0$ is the Fermi surface. The blue band
denotes the thermally broadened shell $|\epsilon|\lesssim T$ in which
electron-hole excitations are available. The black dot represents
an electron excitation above the Fermi surface, and the open circle
represents a hole below it. Thus $\gamma e^{-}\to he^{-}$ is a photon-to-graviton
conversion accompanied by an electron-hole transition near the Fermi
surface.}
\label{fig:photo_fermi_surface}
\end{figure}

Eq.~\eqref{eq:photo_stat_identity} reveals a crucial physical insight:
the reaction is best viewed not as ordinary Compton scattering on
a free electron, but as photon-to-graviton conversion accompanied
by a single electron-hole transition near the Fermi surface, as illustrated
in Fig.~\ref{fig:photo_fermi_surface}. 

Interestingly, Eq.~\eqref{eq:photo_stat_identity} exhibits a symmetry
under simultaneous exchanges of $p_{1}\leftrightarrow k_{1}$ and
$p_{2}\leftrightarrow\omega$. Note that, either $f_{e}(p_{1})[1-f_{e}(k_{1})]$
or $[1-f_{e}(p_{1})]f_{e}(k_{1})$ being suppressed would make this
process very inefficient. 

In the strong degeneracy limit with fixed $\omega$ and $p_{2}$,
it is straightforward to obtain
\begin{equation}
\lim_{T/\mu_{e}\to0}\int dp_{1}[1-f_{e}(p_{1})]f_{e}(k_{1})=\frac{q_{0}e^{q_{0}/T}}{e^{q_{0}/T}-1}\thinspace,\label{eq:-10}
\end{equation}
where $q_{0}\equiv p_{1}-k_{1}=\omega-p_{2}$.  For $q_{0}>0$, this
is of order $q_{0}$, while for $q_{0}<0$ it is exponentially suppressed.
To avoid the exponential suppression, $p_{2}<\omega$ (equivalent
to $q_{0}>0$) is required. Therefore, to generate a graviton with
a fixed energy $\omega$ via this process, an exothermic reaction
($q_{2}<\omega$) is more efficient than an endothermic reaction ($q_{2}>\omega$).

The squared amplitude of $e^{-}+\gamma\to e^{-}+h$ in vacuum contains
a $t$-channel IR divergence, which in thermal plasma is regulated
by Debye-H\"uckel screening. We refer to Ref.~\cite{Xu:2025wjq}
for a detailed calculation of the squared amplitude including the
screening effect and adopt the result obtained therein {[}a factor
of $4$ has been included due to Eq.~\eqref{eq:-7}{]}:
\begin{equation}
|{\cal M}|^{2}=\frac{e^{2}}{M_{P}^{2}}\frac{s^{2}+u^{2}}{\kappa^{2}-t}\thinspace,\label{eq:photo_amp_screened}
\end{equation}
where  the Mandelstam variables are defined as: 
\begin{equation}
\begin{aligned}s & =(p_{1}+p_{2})^{2}=(k_{1}+k)^{2}\thinspace,\\
t & =(p_{1}-k_{1})^{2}=(p_{2}-k)^{2}\thinspace,\\
u & =(p_{1}-k)^{2}=(p_{2}-k_{1})^{2}\thinspace.
\end{aligned}
\end{equation}
The factor $\kappa$ is the Debye-H\"uckel screening scale, determined
by $\kappa^{2}=e^{2}\sum_{i}Q_{i}^{2}n_{i}/T$ with $Q_{i}$ and $n_{i}$
the electric charge and number density of particle $i$.  

Assuming the electron momenta are near the Fermi surface ($p_{1}\sim k_{1}\sim\mu_{e}$)
and the electron directions can be averaged out, we approximate $|{\cal M}|^{2}$
as follows: 
\begin{align}
|{\cal M}|^{2} & \simeq\frac{4\mu_{e}^{2}e^{2}}{M_{P}^{2}}\cdot\frac{p_{2}^{2}(1-\cos\theta_{p_{1},p_{2}})^{2}+\omega^{2}(1-\cos\theta_{p_{1},k})^{2}}{\kappa^{2}+2p_{2}\omega(1-\cos\theta_{p_{2},k})}\nonumber \\
 & \simeq\frac{4\mu_{e}^{2}e^{2}}{M_{P}^{2}}\cdot\frac{p_{2}^{2}+\omega^{2}}{\kappa^{2}+2p_{2}\omega(1-\cos\theta_{p_{2},k})}.\label{eq:M2-approx}
\end{align}

Substituting Eq.~\eqref{eq:M2-approx} into Eq.~\eqref{eq:photo_rate_start},
we proceed with the phase space integration. Since the approximate
form of $|{\cal M}|^{2}$ in Eq.~\eqref{eq:M2-approx} does not depend
on $p_{1}$ and $k_{1}$, we first integrate out $p_{1}$ and $k_{1}$:
\begin{align}
 & \int\frac{d^{3}\mathbf{p}_{1}}{2p_{1}}\frac{d^{3}\mathbf{k}_{1}}{2k_{1}}[1-f_{e}(p_{1})]f_{e}(k_{1})\delta^{(4)}(p_{1}-k_{1}-q)\nonumber \\
= & \int\frac{d^{3}\mathbf{p}_{1}}{4p_{1}(p_{1}-q_{0})}[1-f_{e}(p_{1})]f_{e}(p_{1}-q_{0})\delta^{(1)}\left(p_{1}-q_{0}-\sqrt{p_{1}^{2}+q^{2}-2p_{1}q\cos\theta}\right)\nonumber \\
\simeq & \frac{\pi}{2q}\int dp_{1}[1-f_{e}(p_{1})]f_{e}(p_{1}-q_{0})\nonumber \\
\simeq & \frac{\pi}{2q}\frac{q_{0}e^{q_{0}/T}}{e^{q_{0}/T}-1}\thinspace,\label{eq:-11}
\end{align}
where $q^{\mu}\equiv p_{2}^{\mu}-k^{\mu}=(q_{0},\mathbf{q})$, $q\equiv\left|\mathbf{q}\right|$,
and in the last step we have used Eq.~\eqref{eq:-10}. 

Using Eq.~\eqref{eq:-11}, we rewrite Eq.~\eqref{eq:photo_rate_start}
as 
\begin{align}
\Rate & =\frac{n_{\gamma}n_{e}}{(2\pi)^{6}N_{\gamma e}}\omega e^{-\omega/T}\frac{(2\pi)^{4}}{2}\int\frac{d^{3}p_{2}}{2p_{2}}d\Omega_{h}|{\cal M}|^{2}\frac{\pi}{2q}\frac{q_{0}e^{q_{0}/T}}{e^{q_{0}/T}-1}\thinspace.\label{eq:-12}
\end{align}
The integral in Eq.~\eqref{eq:-12} can be computed as follows:
\begin{align}
I_{\gamma e} & \equiv\int\frac{d^{3}p_{2}}{2p_{2}}d\Omega_{h}|{\cal M}|^{2}\frac{\pi}{2q}\frac{q_{0}e^{q_{0}/T}}{e^{q_{0}/T}-1}\nonumber \\
 & =2\pi^{3}\int dp_{2}dc_{2}\frac{|{\cal M}|^{2}}{q}\frac{q_{0}e^{q_{0}/T}}{e^{q_{0}/T}-1}\nonumber \\
 & =8\pi^{3}\frac{e^{2}\mu_{e}^{2}}{M_{P}^{2}}\int dp_{2}p_{2}(p_{2}^{2}+\omega^{2})\frac{q_{0}e^{q_{0}/T}}{e^{q_{0}/T}-1}\int\frac{dc_{2}}{q[\kappa^{2}+2p_{2}\omega(1-c_{2})]}\thinspace,\label{eq:-13}
\end{align}
where $c_{2}\equiv\cos\theta_{p_{2},k}$. 

Using $q^{2}=p_{2}^{2}+\omega^{2}-2p_{2}\omega c_{2}$, which implies
$q\,dq=-p_{2}\omega\,dc_{2}$, one has 
\begin{equation}
\int\frac{dc_{2}}{q[\kappa^{2}+2p_{2}\omega(1-c_{2})]}=\frac{1}{p_{2}\omega}\int_{|p_{2}-\omega|}^{p_{2}+\omega}\frac{dq}{q^{2}+\kappa^{2}-(p_{2}-\omega)^{2}}\thinspace.\label{eq:photo_q_integral}
\end{equation}
For $p_{2}\to\omega$, the lower limit approaches zero, and the medium
scale $\kappa$ cuts off the singularity. Since the photon energy
approximately lies in a window $p_{2}\in[0,\omega]$, and the integrand
highly peaks around $p_{2}\simeq\omega$, one can approximate the
integral over $p_{2}$ in Eq.~~\eqref{eq:-13} as 
\begin{equation}
\left[p_{2}(p_{2}^{2}+\omega^{2})\frac{q_{0}e^{q_{0}/T}}{e^{q_{0}/T}-1}\right]_{p_{2}=\omega/2}\times\omega=\frac{5}{16}\,\frac{\omega^{5}}{1-e^{-\omega/(2T)}}\,,\label{eq:rectangular_app}
\end{equation}
where $q_{0}=\omega-p_{2}$ has been used. With this approximation,
Eq.~\eqref{eq:-13} gives 
\begin{equation}
I_{\gamma e}\simeq(2\pi)^{3}\frac{e^{2}\mu_{e}^{2}}{M_{P}^{2}}\frac{5}{4}\frac{\omega^{2}}{1-e^{-\omega/(2T)}}{\cal F}\left(\frac{\omega}{\kappa}\right),\label{eq:photo_I_closed}
\end{equation}
where
\begin{equation}
{\cal F}(x)\equiv\begin{cases}
\frac{\theta}{\tan\theta}\ \ \text{with}\ \theta\equiv\arccos(x/2)\ \  & \text{for\ }x\leq2\\
\frac{\theta}{\tanh\theta}\ \text{with}\ \theta\equiv\text{arccosh}(x/2)\ \  & \text{for\ }x>2
\end{cases}\thinspace.\label{eq:-14}
\end{equation}
 In principle, the first case in Eq.~\eqref{eq:-14} can be applied
to $x>2$ as well if analytical continuation is allowed. In this case,
$\theta$ would be a complex value while $\theta/\tan\theta$ is still
real. In practice, we present the two cases separately to avoid potential
confusion. 

Putting the above pieces together, the photoproduction rate is therefore
\begin{align}
\Rate & =\frac{n_{\gamma}n_{e}}{(2\pi)^{6}N_{\gamma e}}\omega e^{-\omega/T}\frac{(2\pi)^{4}}{2}I_{\gamma e}\label{eq:-17}\\
 & \simeq\frac{15\pi\alpha Gn_{\gamma}n_{e}}{4T^{3}\mu_{e}I_{e}}\frac{\omega^{3}e^{-\omega/T}}{1-e^{-\omega/(2T)}}{\cal F}\left(\frac{\omega}{\kappa}\right).\label{eq:-15}
\end{align}
 Here, the result contains $n_{\gamma}n_{e}$ in the numerator but
its dependence on $n_{e}$ is actually not linear because the denominator
contains $\mu_{e}\propto n_{e}^{1/3}$. Using Eq.~\eqref{eq:-6} and
the $N_{\gamma e}$-${\cal N}_{\gamma e}$ relation  in Eq.~\eqref{eq:-8},
we can also write Eq.~\eqref{eq:-17} as
\begin{align}
\frac{d\Gamma}{dVd\omega} & \simeq\frac{5\alpha G}{2\pi^{3}}\mu_{e}^{2}\frac{\omega^{3}e^{-\omega/T}}{1-e^{-\omega/(2T)}}{\cal F}\left(\frac{\omega}{\kappa}\right)\label{eq:-16}\\
 & \simeq\frac{5\alpha G}{2\pi^{3}}\left(3\pi^{2}n_{e}\right)^{\frac{2}{3}}\frac{\omega^{3}e^{-\omega/T}}{1-e^{-\omega/(2T)}}{\cal F}\left(\frac{\omega}{\kappa}\right),\label{eq:-35}
\end{align}
which exhibits an approximate explicit density scaling $n_{e}^{2/3}$
in the strong-degeneracy limit, before accounting for the density
dependence of the screening function. 

 In the upper left  panel of Fig.~\ref{fig:MC-ana}, we compare
the analytical result with the corresponding MC result. The analytical
spectrum is slightly larger than the MC spectrum near the peak. This
difference mainly originates from the rectangular approximation in
Eq.~(\ref{eq:rectangular_app}), where the photon-energy integral
is approximated by the width of the kinematically allowed interval
multiplied by the integrand evaluated at its maximum. Since the integrand
decreases away from the peak, this replacement slightly overestimates
the true integral and therefore leads to a modest enhancement of the
analytical spectrum.


\section{Electron-positron annihilation: $e^{-}+e^{+}\to h+\gamma$}

\label{sec:ee_annihilation}

We denote the momenta of initial and final states of this process
as follows
\begin{equation}
e^{-}(p_{1})+e^{+}(p_{2})\to h(k)+\gamma(q)\thinspace.
\end{equation}

Similar to the calculation in Sec.~\ref{sec:e-gamma}, the production
rate reads 
\begin{equation}
\frac{d\Gamma}{dV\,d\omega}=\frac{n_{e^{-}}n_{e^{+}}\omega}{(2\pi)^{6}N_{-}N_{+}}\int\frac{d^{3}p_{1}\,d^{3}p_{2}\,d^{3}q\,d\Omega_{h}}{16E_{1}E_{2}q}f_{-}(p_{1})f_{+}(p_{2})|\mathcal{M}|^{2}(2\pi)^{4}\delta^{(4)}(p_{1}+p_{2}-k-q)\thinspace,\label{eq:ee_ann_rate_1}
\end{equation}
where $N_{\mp}=\int d^{3}p\,f_{\mp}(p)$, $f_{\pm}(p)=1/\left[\exp\left(\frac{p\pm\mu_{e}}{T}\right)+1\right]$,
and the squared amplitude $|\mathcal{M}|^{2}$ is given by~\cite{Xu:2025wjq}
\begin{equation}
|\mathcal{M}|^{2}=\frac{e^{2}}{M_{P}^{2}}\frac{t^{2}+u^{2}}{s}\thinspace.
\end{equation}

 Integrating over the photon momentum $d^{3}q$ and using $n_{e^{+}}/N_{+}=2/(2\pi)^{3}$,
we have 
\begin{align}
\frac{d\Gamma}{dV\,d\omega} & =\frac{2\omega n_{e^{-}}}{(2\pi)^{9}N_{-}}\int\frac{d^{3}p_{1}\,d^{3}p_{2}\,d^{3}q\,d\Omega_{h}}{4p_{1}p_{2}\,4q}f_{-}(p_{1})f_{+}(p_{2})|\mathcal{M}|^{2}(2\pi)^{4}\delta^{(4)}(p_{1}+p_{2}-k-q)\nonumber \\
 & =\frac{\omega n_{e^{-}}}{8(2\pi)^{5}N_{-}}\int\frac{d^{3}p_{1}\,d^{3}p_{2}}{p_{1}p_{2}}\,d\Omega_{h}\,f_{-}f_{+}|\mathcal{M}|^{2}\delta(E_{12})\nonumber \\
 & =\frac{\omega n_{e^{-}}}{32\pi^{3}N_{-}}\int_{0}^{\infty}dp_{1}\,dp_{2}\,p_{1}p_{2}f_{-}f_{+}\int_{-1}^{1}dx\int_{-1}^{1}dc\int_{0}^{2\pi}d\phi\,|\mathcal{M}|^{2}\delta(E_{12})\thinspace,\label{eq:ee_ann_rate_2}
\end{align}
with
\begin{equation}
E_{12}\equiv p_{1}p_{2}(1-c)-\omega p_{1}(1-x)-\omega p_{2}(1-c_{2h})\thinspace.\label{eq:-18}
\end{equation}
Here we have chose $\mathbf{p}_{1}$ as the polar-axis and defined
$c\equiv\hat{\mathbf{p}}_{1}\cdot\hat{\mathbf{p}}_{2}$, $x\equiv\hat{\mathbf{p}}_{1}\cdot\hat{\mathbf{k}}$,
and $c_{2h}\equiv cx+\sqrt{1-c^{2}}\sqrt{1-x^{2}}\cos\phi$. From
the second to the last step in  Eq.~\eqref{eq:ee_ann_rate_2}, we
have expressed the angular measure as $d\Omega_{p_{1}}\,d\Omega_{h}\,d\Omega_{p_{2}}=(4\pi)(2\pi\,dx)(dc\,d\phi)=8\pi^{2}\,dx\,dc\,d\phi.$
Since $\int dx\,\delta(E_{12})\simeq(\omega p_{1})^{-1}$, Eq.~(\ref{eq:ee_ann_rate_2})
simplifies to 
\begin{align}
\frac{d\Gamma}{dV\,d\omega}\simeq\frac{n_{e^{-}}}{16\pi^{2}N_{-}}\int_{0}^{\infty}dp_{1}\int_{0}^{\infty}dp_{2}\,p_{2}f_{-}(p_{1})f_{+}(p_{2})\int dc\,|\mathcal{M}|^{2}.\label{eq:ee_ann_rate_3}
\end{align}

To proceed further, we need to express $|\mathcal{M}|^{2}$ as a function
of $p_{1},p_{2}$ and $c$. To this end, we write $s=(p_{1}+p_{2})^{2}=2p_{1}p_{2}(1-c)$,
$t=(p_{1}-k)^{2}=-2p_{1}\omega(1-x)$, and $u=(p_{2}-k)^{2}=-2p_{2}\omega(1-c_{2h})$.
For degenerate electron gas, the relevant hierarchy is $p_{1}\sim\mu\gg p_{2}\sim T$.
To the leading order in $p_{2}/p_{1}$, the energy constraint with
$E_{12}=0$ gives $t\simeq-s(1-\omega/p_{1})$, $u\simeq-s\omega/p_{1}$,
and hence 
\begin{equation}
|\mathcal{M}|^{2}\simeq\frac{2e^{2}}{M_{P}^{2}}p_{1}p_{2}(1-c)\left[\left(1-\frac{\omega}{p_{1}}\right)^{2}+\left(\frac{\omega}{p_{1}}\right)^{2}\right].\label{eq:ee_amplitude_approx}
\end{equation}
Energy conservation $E_{12}=0$ further gives $p_{2}(1-c)\simeq\omega(1-x)$.
Since $0\leq1-x\leq2$, the available initial-state angular range
is restricted to $0\leq1-c\lesssim2\omega/p_{2}$ for $\omega<p_{2}$.
On the other hand, $0\leq1-c\lesssim2$ for $\omega>p_{2}$. Together
with the factor $(1-c)$ in the squared amplitude, this gives 
\begin{equation}
\int dc\,(1-c)\simeq2S_{{\rm IR}}(p_{2},\omega),\qquad S_{{\rm IR}}(p_{2},\omega)\equiv\min\left[1,\frac{\omega^{2}}{p_{2}^{2}}\right].\label{eq:ee_IR}
\end{equation}
Substituting Eq.~(\ref{eq:ee_IR}) and the squared amplitude into
Eq.~(\ref{eq:ee_ann_rate_3}) gives 
\begin{equation}
\frac{d\Gamma}{dV\,d\omega}\simeq\frac{e^{2}n_{e^{-}}}{4\pi^{2}M_{P}^{2}N_{-}}\,I_{-}(\omega)I_{+}(\omega)\thinspace,\label{eq:ee_ann_rate_4}
\end{equation}
where we have defined
\begin{align}
I_{-}(\omega) & \equiv\int_{\omega}^{\infty}dp_{1}\,p_{1}f_{-}(p_{1})\left[\left(1-\frac{\omega}{p_{1}}\right)^{2}+\left(\frac{\omega}{p_{1}}\right)^{2}\right],\\
I_{+}(\omega) & \equiv\int_{0}^{\infty}dp_{2}\,p_{2}^{2}f_{+}(p_{2})S_{{\rm IR}}(p_{2},\omega)\thinspace.\label{eq:ee_integrals}
\end{align}

In the strongly degenerate regime, $\mu_{e}\gg T$, the positron distribution
is well approximated by $f_{+}(p_{2})\simeq e^{-(p_{2}+\mu_{e})/T}$.
Since $S_{{\rm IR}}=1$ for $p_{2}<\omega$ and $S_{{\rm IR}}=\omega^{2}/p_{2}^{2}$
for $p_{2}>\omega$, the positron integral can be evaluated directly,
\begin{align}
I_{+}(\omega) & \simeq e^{-\mu_{e}/T}\left[\int_{0}^{\omega}dp_{2}\,p_{2}^{2}e^{-p_{2}/T}+\omega^{2}\int_{\omega}^{\infty}dp_{2}\,e^{-p_{2}/T}\right]\nonumber \\
 & =2T^{3}e^{-\mu_{e}/T}\left[1-\left(1+\frac{\omega}{T}\right)e^{-\omega/T}\right].\label{eq:Iplus}
\end{align}

We now compute $I_{-}(\omega)$.  Since $\mu_{e}\gg T$, we have
$f_{-}(p_{1})\to1$ for $p_{1}\leq\mu_{e}$. This corresponds to a
zero-temperature contribution from electrons below the Fermi surface
\begin{align}
I_{-}^{0}(\omega) & =\int_{\omega}^{\mu_{e}}dp_{1}\left(p_{1}-2\omega+\frac{2\omega^{2}}{p_{1}}\right)\nonumber \\
 & =\frac{1}{2}\left(\mu_{e}^{2}-4\mu_{e}\omega+3\omega^{2}+4\omega^{2}\ln\frac{\mu_{e}}{\omega}\right)\Theta(\mu_{e}-\omega)\equiv\frac{1}{2}H(\mu_{e},\omega)\Theta(\mu_{e}-\omega)\,,\label{eq:0T}
\end{align}
which is valid for soft gravitons. For hard graviton production from
electrons above the Fermi surface with $p\geq\mu_{e}$, we must keep
the exact form of $f_{-}(p)$. This leads to
\begin{align}
I_{-}^{T}(\omega) & =\int_{\omega}^{\infty}dp\frac{1}{e^{(p-\mu_{e})/T}+1}\left(p-2\omega+\frac{2\omega^{2}}{p}\right)\label{eq:finiteT-1}
\end{align}
which is not elementary because of the last term in parentheses. Writing
$p=\omega+\delta$, where $\delta$ is assumed to be small for hard
gravitons, the kinematic factor has the controlled expansion $p-2\omega+\frac{2\omega^{2}}{p}=\omega-\delta+2\delta^{2}/\omega+\mathcal{O}\left(\delta^{3}/\omega^{2}\right)$.
Keeping the leading term therefore gives the finite-temperature contribution
\begin{align}
I_{-}^{T}(\omega) & \simeq\omega\int_{\omega}^{\infty}\frac{dp}{e^{(p-\mu)/T}+1}=\omega T\ln\left(1+e^{(\mu_{e}-\omega)/T}\right)\,,\label{eq:finiteT}
\end{align}
which is suppressed for high frequency gravitons. In the zero-temperature
limit, we note that $I_{-}^{T\to0}(\omega)\to\omega(\mu_{e}-\omega)\Theta(\mu_{e}-\omega)$,
which matches with Eq.~(\ref{eq:0T}) at the edge $\mu_{e}=\omega$.
Combining both contributions, we have the following compact approximation
\begin{equation}
I_{-}(\omega)\simeq\frac{1}{2}\widetilde{H}_{T}(\mu_{e},\omega)\equiv\frac{1}{2}\left\{ H(\mu_{e},\omega)\Theta(\mu_{e}-\omega)+2\omega\left[T\ln\left(1+e^{(\mu_{e}-\omega)/T}\right)-(\mu_{e}-\omega)\Theta(\mu_{e}-\omega)\right]\right\} \,.
\end{equation}

Using  $N_{-}\simeq4\pi\mu_{e}^{3}I_{e}/3$, together with $e^{2}=4\pi\alpha$
and $1/M_{P}^{2}=8\pi G$, we finally obtain 
\begin{equation}
\Rate\simeq\frac{6\alpha Gn_{e^{-}}}{\pi\mu_{e}^{3}I_{e}}T^{3}e^{-\mu_{e}/T}B\left(\frac{\omega}{T}\right)\widetilde{H}_{T}(\mu_{e},\omega)\thinspace,\label{eq:final}
\end{equation}
where 
\begin{align}
\widetilde{H}_{T}(\mu_{e},\omega) & \equiv H(\mu_{e},\omega)\Theta(\mu_{e}-\omega)+2\omega\left[T\ln\left(1+e^{(\mu_{e}-\omega)/T}\right)-(\mu_{e}-\omega)\Theta(\mu_{e}-\omega)\right],\label{eq:-24}\\
H(\mu_{e},\omega) & \equiv\mu_{e}^{2}-4\mu_{e}\omega+3\omega^{2}+4\omega^{2}\ln\frac{\mu_{e}}{\omega}\thinspace,\label{eq:-25}\\
B(x) & \equiv1-(1+x)e^{-x}\thinspace.\label{eq:-26}
\end{align}

The upper right panel of Fig.~\ref{fig:MC-ana} compares the above
analytical result with the corresponding MC result, showing modest
agreement.



\section{Bremsstrahlung: $e^{-}e^{-}\to e^{-}e^{-}h$ }

We denote the momenta of initial and final states of this process
as follows
\begin{equation}
e^{-}(p_{1})+e^{-}(p_{2})\longrightarrow e^{-}(k_{1})+e^{-}(k_{2})+h(k)\thinspace.\label{eq:ee-process}
\end{equation}

This process contains identical electrons in the initial and final
states. Hence the following symmetry factor should be taken into account:
\begin{equation}
S_{{\rm stat}}^{ee}=\frac{1}{2!_{{\rm in}}\,2!_{{\rm out}}}=\frac{1}{4}\thinspace.\label{eq:ee-stat}
\end{equation}
In our analytical convention this symmetry factor is included in the
squared matrix element.  

The graviton production rate for this process reads
\begin{align}
\Rate= & \frac{(2\pi)^{6}n_{e}^{2}}{N_{e}^{2}}\int\frac{\dd^{3}p_{1}}{(2\pi)^{3}2p_{1}}\frac{\dd^{3}p_{2}}{(2\pi)^{3}2p_{2}}\frac{\dd^{3}k_{1}}{(2\pi)^{3}2k_{1}}\frac{\dd^{3}k_{2}}{(2\pi)^{3}2k_{2}}\frac{\omega\,\dd\Omega_{h}}{2(2\pi)^{3}}(2\pi)^{4}\delta^{(4)}(p_{1}+p_{2}-k_{1}-k_{2}-k)\nonumber \\
 & \times f_{e}(p_{1})f_{e}(p_{2})[1-f_{e}(k_{1})][1-f_{e}(k_{2})]|\mathcal{M}|^{2}\thinspace,\label{eq:ee-full-rate}
\end{align}
where $N_{e}\equiv4\pi\int_{0}^{\infty}\dd p\,p^{2}f_{e}(p)$. 

The process $e^-e^-\to e^-e^-h$ receives 14 tree-level contributions, including both $t$- and $u$-channel topologies in which the graviton is emitted from an external electron leg, an interaction vertex, or the exchanged photon. When the graviton energy is small compared with the characteristic energy scale, Weinberg’s soft graviton theorem~\cite{Weinberg:1965nx} implies that the $2\to3$ amplitude factorizes into the $2\to2$ amplitude and a universal soft factor constructed from the scalar products of four  momenta of graviton and external particles, i.e., $k_i \cdot k$ and $p_i \cdot k$ with $i=1,2.$
The squared matrix element $|{\cal M}|^{2}$ at the leading soft order
can then be obtained using the soft graviton theorem.
Including Debye screening and the symmetry factor in Eq.~\eqref{eq:ee-stat},
the result reads
\begin{equation}
|\mathcal{M}|^{2}\simeq    e^{4}\frac{(s^{2}+st+t^{2})^{2}}{tu(\kappa^{2}-t)(\kappa^{2}-u)}\,S_{g}\thinspace,\label{eq:-19}
\end{equation}
where $s$, $t$, $u$ are the Mandelstam variables of the corresponding
M{\o}ller scattering process ($e^{-}e^{-}\to e^{-}e^{-}$), and the
soft radiation factor $S_{g}$ is 
\begin{equation}
S_{g}=\frac{2}{\MP^{2}}\left[\frac{s^{2}}{XY}+\frac{s^{2}}{AB}-\frac{t^{2}}{XA}-\frac{u^{2}}{XB}-\frac{u^{2}}{YA}-\frac{t^{2}}{YB}\right],\label{eq:Sg-cov-1}
\end{equation}
with $X\equiv2p_{1}\cdot k$, $Y\equiv2p_{2}\cdot k$, $A\equiv2k_{1}\cdot k$,
$B\equiv2k_{2}\cdot k$. Each of $X,Y,A,B$ scales as $\mu\omega$,
and therefore $S_{g}\sim\omega^{-2}$.

By defining 
\begin{equation}
\epsilon_{1,2,3,4}\equiv(p_{1},\ p_{2},\ k_{1},\ k_{2})-\mu_{e}\label{eq:-20}
\end{equation}
and assuming $|\epsilon_{i}|\sim\mathcal{O}(T)\ll\mu_{e}$, we approximate
the three-momentum part of $\delta^{(4)}$ in Eq.~\eqref{eq:ee-full-rate}
as $\delta^{(3)}(\bm{p}_{1}+\bm{p}_{2}-\bm{k}_{1}-\bm{k}_{2})$ and
write the energy part as $\delta(\epsilon_{1}+\epsilon_{2}-\epsilon_{3}-\epsilon_{4}-\omega)$.
Then the product rate can be approximately factorized as
\begin{equation}
\Rate\simeq\frac{n_{e}^{2}\mu^{4}\omega}{N_{e}^{2}(4\pi)^{5}}I_{ee}(\omega,T)\,\mathcal{A}_{ee}(\omega)\thinspace,\label{eq:ee-rate-factorized}
\end{equation}
where
\begin{align}
I_{ee}(\omega,T) & \equiv\int_{-\infty}^{\infty}\dd\epsilon_{1}\dd\epsilon_{2}\dd\epsilon_{3}\dd\epsilon_{4}\;f_{e}(\epsilon_{1})f_{e}(\epsilon_{2})[1-f_{e}(\epsilon_{3})][1-f_{e}(\epsilon_{4})]\delta(\epsilon_{1}+\epsilon_{2}-\epsilon_{3}-\epsilon_{4}-\omega)\thinspace,\label{eq:Iee-def}\\
\mathcal{A}_{ee}(\omega) & \equiv\int\dd\Omega_{p_{1}}\dd\Omega_{p_{2}}\dd\Omega_{k_{1}}\dd\Omega_{k_{2}}\dd\Omega_{h}\delta^{(3)}\left(\bm{p}_{1}+\bm{p}_{2}-\bm{k}_{1}-\bm{k}_{2}\right)|\mathcal{M}|^{2}\thinspace.\label{eq:Aee-def}
\end{align}

The integral $I_{ee}$ can be viewed as a four-electron generalization
of Eq.~\eqref{eq:-10} or \eqref{eq:-11}. With similar approximations,
we obtain
\begin{equation}
I_{ee}(\omega,T)=\frac{\omega(4\pi^{2}T^{2}+\omega^{2})}{6(e^{\omega/T}-1)}.\label{eq:Iee-final}
\end{equation}

The angular integral $\mathcal{A}_{ee}$ can be approximately reduced
to
\begin{equation}
\mathcal{A}_{ee}(\omega)\simeq\frac{512\pi^{3}e^{4}}{\mu_{e}\MP^{2}\omega^{2}}\mathcal{K}_{ee}\left(\frac{\kappa^{2}}{2\mu_{e}^{2}}\right),\label{eq:ee-Aee-compact}
\end{equation}
with
\begin{equation}
\mathcal{K}_{ee}(x)\simeq\frac{39}{4}-\frac{7}{2}\ln x+\frac{8}{3}\left(\ln x\right)^{2}\thinspace.\label{eq:ee-K-final}
\end{equation}

Putting the above pieces together and using  $N_{e}\simeq(4\pi/3)\mu_{e}^{3}I_{e}$,
we obtain
\begin{align}
\Rate & \simeq\frac{6\alpha^{2}Gn_{e}^{2}}{\pi\mu_{e}^{3}I_{e}^{2}}\frac{4\pi^{2}T^{2}+\omega^{2}}{e^{\omega/T}-1}\mathcal{K}_{ee}\left(\frac{\kappa^{2}}{2\mu_{e}^{2}}\right) \nonumber \\
 & \simeq\frac{2\alpha^{2}G\mu_{e}^{3}}{3\pi^{5}}\thinspace\frac{4\pi^{2}T^{2}+\omega^{2}}{e^{\omega/T}-1}\mathcal{K}_{ee}\left(\frac{\kappa^{2}}{2\mu_{e}^{2}}\right) \label{eq:ee-rate-closed}.
\end{align}
In the lower right panel of Fig.~\ref{fig:MC-ana}, we  compare the above analytical result with the corresponding MC computation, and find that they agree well with each other. For cross checking,  we have also computed explicitly the matrix elements $|\mathcal{M}|^{2} $ for $2\to 3$ including the 14 production without invoking the soft graviton theorem, and found that Eq.~\eqref{eq:ee-rate-closed} remains a good approximation.


\section{Bremsstrahlung: $e^{-}p\to e^{-}p\,h$}

Since protons are approximately nonrelativistic in a supernova core,
this process can be treated as electron scattering off a static Coulomb
potential. To extend the applicability of our calculation, we consider
a generic heavy scatterer with electric charge $Z$ rather than restricting
ourselves to a proton, and denote the momenta of initial and final
states of the $e^{-}Z$ scattering process as follows:
\begin{equation}
e^{-}(p)+Z(P)\longrightarrow e^{-}(p')+Z(P')+h(k)\thinspace.\label{eq:eZ-process}
\end{equation}
It is convenient to establish the scattering kinematics in the frame
where $Z$ is initially at rest, such that
\begin{equation}
P=(M,\bm{0}),\qquad p=E(1,\hat{\bm{u}}_{i}),\qquad p'=E_{f}(1,\hat{\bm{u}}_{f}),\qquad k=\omega(1,\hat{\bm{n}}),
\end{equation}
where $M$ is the mass of the heavy scatterer, and $\hat{\bm{u}}_{i,f}$
and $\hat{\bm{n}}$ are unit vectors. We work in the regime $M\gg\mu_{e}\gg T$
and use the leading-soft expansion $\omega/E\ll1$, with $E\sim\mu_{e}$. 

Imposing the on-shell condition $P'^{2}=M^{2}$ with $P'=P+p-p'-k$
gives
\begin{equation}
2M(E-E_{f}-\omega)-2EE_{f}(1-c_{e})-2E\omega(1-c_{i})+2E_{f}\omega(1-c_{f})=0\thinspace,\label{eq:-21}
\end{equation}
with 
\begin{equation}
c_{i}=\hat{\bm{u}}_{i}\cdot\hat{\bm{n}}\thinspace,\qquad c_{f}=\hat{\bm{u}}_{f}\cdot\hat{\bm{n}}\thinspace,\qquad c_{e}=\hat{\bm{u}}_{i}\cdot\hat{\bm{u}}_{f}\thinspace.
\end{equation}
Eq.~\eqref{eq:-21} can be solved for the final electron energy: 
\begin{equation}
E_{f}=\frac{M(E-\omega)-E\omega(1-c_{i})}{M+E(1-c_{e})-\omega(1-c_{f})}.\label{eq:Ef-exact}
\end{equation}
The spatial momentum transferred to the target is $\bm{Q}=\bm{p}-\bm{p}'-\bm{k}$,
with 
\begin{equation}
Q^{2}\equiv|\bm{Q}|^{2}=E^{2}+E_{f}^{2}+\omega^{2}-2EE_{f}c_{e}-2E\omega c_{i}+2E_{f}\omega c_{f}\thinspace.\label{eq:eZ-Q-exact}
\end{equation}

We take the limits in the order 
\begin{equation}
\frac{\omega}{E}\longrightarrow0\quad\text{at fixed }M\thinspace,\qquad\frac{E}{M}\longrightarrow0\quad\text{afterwards}\thinspace.\label{eq:eZ-order}
\end{equation}
At finite $M$, the first limit retains ordinary elastic recoil, $E_{f}\to ME/[M+E(1-c_{e})]$.
The subsequent large-$M$ limit gives $E_{f}\simeq E$ in the hard
scattering kernel and 
\begin{equation}
Q^{2}\simeq2E^{2}(1-c_{e})\thinspace.\label{eq:eZ-Q-soft}
\end{equation}
The energy difference $E-E_{f}=\omega$ is nevertheless kept in the
thermal phase space.

The graviton production rate for this process reads
\begin{align}
\Rate= & \frac{(2\pi)^{6}n_{e}n_{Z}}{N_{e}N_{Z}}\int\frac{\dd^{3}p}{(2\pi)^{3}2E}\frac{\dd^{3}P}{(2\pi)^{3}2M}\frac{\dd^{3}p'}{(2\pi)^{3}2E_{f}}\frac{\dd^{3}P'}{(2\pi)^{3}2M}\nonumber \\
 & \times\frac{\omega\,\dd\Omega_{h}}{2(2\pi)^{3}}(2\pi)^{4}\delta^{(4)}(p+P-p'-P'-k)f_{e}(E)f_{Z}(P)[1-f_{e}(E_{f})]|\mathcal{M}|^{2},\label{eq:eZ-full-rate}
\end{align}
where $N_{e}\equiv\int\dd^{3}Pf_{e}(p)$ and $N_{Z}\equiv\int\dd^{3}P\,f_{Z}(P)$.
The thermal distribution of $Z$ can be treated as Maxwell--Boltzmann
at the accuracy considered here. 

In the large-$M$ limit, the electromagnetic current of the scatterer
reduces to $\bar{u}(P')\gamma^{\alpha}u(P)\simeq2Mv^{\alpha}$ with
$v^{\alpha}=(1,\bm{0})$. The spin-averaged matrix element of $e^{-}+Z\to e^{-}+Z$
(Coulomb scattering) is 
\begin{equation}
|\mathcal{M}_{C}|^{2}=\frac{8Z^{2}e^{4}M^{2}EE_{f}(1+c_{e})}{Q^{4}}\thinspace.\label{eq:Coulomb-kernel}
\end{equation}

Similar to the previous section, the leading-order soft amplitude
of $e^{-}+Z\to e^{-}+Z+h$ factorizes into the above Coulomb scattering
part and a soft radiation factor $S_{g}$:
\begin{equation}
|\mathcal{M}|^{2}=|\mathcal{M}_{C}|^{2}S_{g}\thinspace,\label{eq:-22}
\end{equation}
with
\begin{equation}
S_{g}=\frac{[E(1+c_{i})-E_{f}(1+c_{f})]^{2}+4EE_{f}(1+c_{i})(1+c_{f})\sin^{2}\phi_{eZ}}{2\MP^{2}\omega^{2}}\thinspace.
\end{equation}
Here $\phi_{eZ}$ is the relative azimuth of the initial and final
electron directions around the graviton axis, so that $c_{e}=c_{i}c_{f}+\sqrt{1-c_{i}^{2}}\sqrt{1-c_{f}^{2}}\cos\phi_{eZ}$. 

Including the screening effect by replacing $Q^{-4}\to(Q^{2}+\kappa^{2})^{-2}$
and using $E_{f}\simeq E\simeq\mu$, and Eq.~(\ref{eq:eZ-Q-soft}),
we obtain
\begin{equation}
|\mathcal{M}|^{2}\simeq\frac{4Z^{2}e^{4}M^{2}\mu^{4}(1+c_{e})}{\MP^{2}\omega^{2}[2\mu_{e}^{2}(1-c_{e})+\kappa^{2}]^{2}}\left[(c_{i}-c_{f})^{2}+4(1+c_{i})(1+c_{f})\sin^{2}\phi_{eZ}\right].\label{eq:eZ-screened-degenerate}
\end{equation}

Integrating out $\dd^{3}P'$ with the spatial delta function and $\dd^{3}P$
with $f_{Z}$ (the remaining part of the integral is approximately
independent of $\bm{P}$ in the non-relativistic limit) yields 
\begin{equation}
\Rate\simeq\frac{n_{e}n_{Z}\omega}{(4\pi)^{5}M^{2}N_{e}}\int\frac{\dd^{3}p}{E}\frac{\dd^{3}p'}{E_{f}}\dd\Omega_{h}\,f_{e}(E)[1-f_{e}(E_{f})]\delta(E-E_{f}-\omega)|\mathcal{M}|^{2}\thinspace.\label{eq:eZ-rate-heavy}
\end{equation}
Since $\dd^{3}p/E=E\,\dd E\,\dd\Omega_{p}$ for a massless electron,
the radial and angular integrations can be separated to the leading
order in $T/\mu$ and $\omega/\mu$. Therefore, we factorize the rate
as follows:
\begin{equation}
\Rate_{eZ}(\omega)\simeq\frac{n_{e}n_{Z}\omega}{(4\pi)^{5}M^{2}N_{e}}J_{eZ}(\omega,T)\,\mathcal{A}_{eZ}(\omega),\label{eq:eZ-rate-factorized}
\end{equation}
where 
\begin{align}
J_{eZ}(\omega,T) & =\int_{0}^{\infty}E\,\dd E\int_{0}^{\infty}E_{f}\,\dd E_{f}\,f_{e}(E)[1-f_{e}(E_{f})]\delta(E-E_{f}-\omega)\thinspace,\label{eq:JeZ-def}\\
\mathcal{A}_{eZ}(\omega) & =\int\dd\Omega_{p}\,\dd\Omega_{p'}\,\dd\Omega_{h}|\mathcal{M}|^{2}\thinspace.\label{eq:AeZ-def}
\end{align}

To evaluate $J_{eZ}$, we first set $E_{f}=E-\omega$ according to
the delta function. Near the Fermi surface, $EE_{f}\simeq\mu^{2}$.
Introducing $\epsilon\equiv E-\mu$, we approximate $J_{eZ}$ as 
\begin{align}
J_{eZ}(\omega,T) & \simeq\mu^{2}\int_{-\infty}^{\infty}\dd\epsilon\,f(\epsilon)[1-f(\epsilon-\omega)]\nonumber \\
 & \simeq\mu^{2}\frac{\omega}{e^{\omega/T}-1}\thinspace,\label{eq:JeZ}
\end{align}
where $f(\epsilon)=\frac{1}{e^{\epsilon/T}+1}$. 

Our approximate evaluation of $\mathcal{A}_{eZ}$ in the soft limit
yields
\begin{equation}
\mathcal{A}_{eZ}(\omega)\simeq\frac{64\pi^{3}Z^{2}e^{4}M^{2}}{\MP^{2}\omega^{2}}\mathcal{K}_{eZ}\left(\frac{\kappa^{2}}{2\mu_{e}^{2}}\right),\label{eq:eZ-A-factorized}
\end{equation}
where 
\begin{align}
\mathcal{K}_{eZ}(x) & =-4(1+x)\Li_{2}\!\left(-\frac{2}{x}\right)-\frac{4x+10}{3}\ln\left(1+\frac{2}{x}\right)-\frac{16}{3}\label{eq:LeZ-polylog}\\
 & \simeq2\ln^{2}\left(\frac{2}{x}\right)-\frac{10}{3}\ln\left(\frac{2}{x}\right)+\frac{2\pi^{2}}{3}-\frac{16}{3}\thinspace.
\end{align}
In the second step above, we have assumed $x=\frac{\kappa^{2}}{2\mu_{e}^{2}}\ll1$
and used $\Li_{2}(-2/x)=-\frac{1}{2}\ln^{2}\left(\frac{2}{x}\right)-\frac{\pi^{2}}{6}+\mathcal{O}(x)$.

Assembling the above pieces yields 
\begin{align}
\Rate & \simeq\frac{6Z^{2}\alpha^{2}G\,n_{e}n_{Z}}{\mu_{e}I_{e}}\frac{1}{e^{\omega/T}-1}\mathcal{K}_{eZ}\left(\frac{\kappa^{2}}{2\mu_{e}^{2}}\right)\\
 & \simeq\frac{2Z^{2}\alpha^{2}G\,\mu_{e}^{2}n_{Z}}{\pi^{2}}\frac{1}{e^{\omega/T}-1}\mathcal{K}_{eZ}\left(\frac{\kappa^{2}}{2\mu_{e}^{2}}\right).
 \label{eq:ep}
\end{align}
The lower left panel of Fig.~\ref{fig:MC-ana} compares the above
analytical result with the corresponding MC result, showing good
agreement.


\end{document}